\documentclass[11pt]{article}

\usepackage[top=1in,bottom=1in,left=1in,right=1in]{geometry}
\usepackage{natbib}
\usepackage[unicode=true,bookmarks=true,bookmarksnumbered=false,bookmarksopen=false,breaklinks=false,pdfborder={0 0 1},backref=false,colorlinks=true]{hyperref}
\hypersetup{citecolor=blue}
\usepackage{url}

\usepackage{amsmath}
\usepackage{mathrsfs}
\usepackage{amssymb}
\usepackage{amsthm}
\usepackage[normalem]{ulem}

\usepackage{mathtools}

\usepackage{booktabs}

\newtheorem{theorem}{Theorem}

\usepackage{float}
\usepackage{rotfloat}

\floatstyle{ruled}
\newfloat{algorithm}{tpb}{loa}
\providecommand{\algorithmname}{Algorithm}
\floatname{algorithm}{\protect\algorithmname}

\usepackage{algpseudocode,algorithm}
\algnewcommand\algorithmicinput{\textbf{Input}:}
\algnewcommand\algorithmicoutput{\textbf{Output}:}
\algnewcommand\INPUT{\item[\algorithmicinput]}
\algnewcommand\OUTPUT{\item[\algorithmicoutput]}

\usepackage{graphicx}
\usepackage[caption=false,labelformat=simple]{subfig}

\usepackage{array}
\newcolumntype{L}[1]{>{\raggedright\let\newline\\\arraybackslash\hspace{0pt}}m{#1}}
\newcolumntype{C}[1]{>{\centering\let\newline\\\arraybackslash\hspace{0pt}}m{#1}}
\newcolumntype{R}[1]{>{\raggedleft\let\newline\\\arraybackslash\hspace{0pt}}m{#1}}

\usepackage{rotating}
\usepackage{multirow}

\usepackage{tikz}
\usetikzlibrary{shapes,arrows,backgrounds,calc,positioning,fit,petri,plotmarks}
\usetikzlibrary{arrows}

\usepackage{enumerate}
\usepackage{paralist}

\newcommand*{\affaddr}[1]{#1} 
\newcommand*{\affmark}[1][*]{\textsuperscript{#1}}

\global\long\def\bX{\mathbf{X}}
\global\long\def\bx{\mathbf{x}}
\global\long\def\bY{\mathbf{Y}}

\global\long\def\bZ{\mathbf{Z}}
\global\long\def\bz{\mathbf{z}}

\global\long\def\bD{\mathbf{D}}
\global\long\def\bA{\mathbf{A}}
\global\long\def\bB{\mathbf{B}}
\global\long\def\bC{\mathbf{C}}
\global\long\def\bH{\mathbf{H}}

\global\long\def\bU{\mathbf{U}}

\global\long\def\bV{\mathbf{V}}

\global\long\def\bQ{\mathbf{Q}}
\global\long\def\bR{\mathbf{R}}
\global\long\def\bM{\mathbf{M}}
\global\long\def\bS{\mathbf{S}}

\global\long\def\bT{\mathbf{T}}

\global\long\def\bO{\mathbf{O}}

\global\long\def\balpha{\boldsymbol{\alpha}}

\global\long\def\bvarepsilon{\boldsymbol{\varepsilon}}
\global\long\def\bSigma{\boldsymbol{\Sigma}}
\global\long\def\bgamma{\boldsymbol{\gamma}}
\global\long\def\bdeta{\boldsymbol{\eta}}
\global\long\def\btheta{\boldsymbol{\theta}}
\global\long\def\bphi{\boldsymbol{\phi}}

\global\long\def\bXi{\boldsymbol{\Xi}}

\global\long\def\bTheta{\boldsymbol{\Theta}}

\global\long\def\bzeta{\boldsymbol{\zeta}}

\global\long\def\bzero{\boldsymbol{\mathrm{0}}}

\global\long\def\matI{\boldsymbol{\mathrm{I}}}

\global\long\def\Var{\mathrm{Var}}
\global\long\def\diag{\mathrm{diag}}
\global\long\def\sign{\mathrm{sign}}

\global\long\def\matzero{\ensuremath{\boldsymbol{\mathrm{0}}}}
\global\long\def\matI{\ensuremath{\boldsymbol{\mathrm{I}}}}

\newcommand{\indep}{\rotatebox[origin=c]{90}{$\models$}}

\makeatletter
\newcommand*{\rom}[1]{\expandafter\@slowromancap\romannumeral #1@}
\makeatother

\usepackage{xr}
\makeatletter
\newcommand*{\addFileDependency}[1]{
  \typeout{(#1)}
  \@addtofilelist{#1}
  \IfFileExists{#1}{}{\typeout{No file #1.}}
}
\makeatother
 
\newcommand*{\myexternaldocument}[1]{%
    \externaldocument{#1}%
    \addFileDependency{#1.tex}%
    \addFileDependency{#1.aux}%
}

\myexternaldocument{}

\title{Estimation of Heterogeneous Causal Mediation Effects in a Hypertension Treatment Trial}

\author{%
    Yi Zhao\affmark[1]\thanks{Yi Zhao and Chengyun Li are co-first authors.}, Chengyun Li\affmark[1]\footnotemark[1] and Wanzhu Tu\affmark[1]\thanks{For correspondence, contact Wanzhu Tu at \texttt{wtu1@iu.edu}. This research is supported by National Institutes of Health Grants R01 HL095085 and U24 AA026969.}
    \\
    \affaddr{\affmark[1]Department of Biostatistics and Health Data Science, Indiana University School of Medicine} \\
}

\date{}

\providecommand{\keywords}[1]
{
  {\small 
  \textbf{Keywords:} #1 Causal mediation analysis; Modified covariate estimation; Heterogeneous causal effects; Regularized regression; Precision medicine.}
}

\begin{document}

\maketitle

\thispagestyle{empty}

\begin{abstract}
Hypertension is a highly prevalent condition and a major risk factor for cardiovascular disease. The landmark Systolic Blood Pressure Intervention Trial (SPRINT) showed that lowering systolic blood pressure (BP) goals from $140$ mmHg to $120$ mmHg leads to significantly reduced BP, cardiovascular mortality, and morbidity. However, the underlying mechanisms are not yet fully elucidated. In patients with impaired renal function, early reduction of albuminuria has been proposed as a potential mediation pathway. Evidence from the standard causal mediation analysis (CMA), however, yields inconsistent results, possibly due to heterogeneous mediation effects across individuals. To disseminate the heterogeneity, a new framework that incorporates covariate-treatment and mediator-treatment interactions within a linear structural equation modeling system is introduced. Causal assumptions are discussed and heterogeneous natural direct and indirect effects are parameterized as functions of patient characteristics. A modified covariate approach is proposed to relax the hierarchical constraints and the generalized lasso regularization is employed to ensure parsimony in high-dimensional settings. Asymptotic properties are studied. Simulation studies demonstrate good estimation and inference performance. Analysis of the SPRINT data reveals substantial heterogeneity in mediation effects, identifying a subset of patients who stand to gain from therapies targeting albuminuria.
\end{abstract}
\keywords{}



\clearpage
\setcounter{page}{1}

\section{Introduction}
\label{s:intro}

\subsection{Hypertension and the kidney}

Hypertension (HTN), characterized by persistent elevation of blood pressure (BP), is a highly prevalent condition, especially among older adults~\citep{kearney2005global}. Uncontrolled HTN hastens the onset of complications, including cardio- and cerebrovascular conditions, as well as chronic kidney disease~\cite[CKD,][]{lawes2008global}. While the precise pathophysiology of HTN remains the subject of mechanistic investigations, overwhelming evidence points to sodium retention by the kidney as the main culprit for the increased risk~\citep{o2004salt,he2009comprehensive,mozaffarian2014global}. As the central organ controlling sodium reabsorption, the kidney plays an essential role in regulating BP. Arthur Guyton's classic renal model has demonstrated that HTN arises when the kidney requires greater arterial pressure to excrete sodium to match dietary intake~\citep{guyton1972arterial}. In this pressure-natriuresis process,  an increase in BP can be viewed as a byproduct of the kidney's effort in maintaining sodium balance. When there is a derangement in the kidney's handling of sodium and potassium, the resulting fluid volume expansion contributes to increased BP and gives rise to essential HTN~\citep{adrogue2007sodium}. In a way, the Guytonian model not only outlines the central etiology of HTN, but also underscores why HTN and CKD frequently coexist, each exacerbating the other in a self-reinforcing cycle.

Major classes of anti-HTN therapies have focused on renal mechanisms that promote sodium excretion and fluid reduction. Among these, thiazide and loop diuretics reduce extracellular fluid volume by inhibiting sodium reabsorption in the distal and ascending limbs of the nephron, respectively~\citep{ellison2009thiazide}. A recent clinical trial showed that thiazide-like diuretic chlorthalidone has a superb BP-lowering effect in patients with CKD~\citep{agarwal2021chlorthalidone}. Calcium channel blockers (CCB) and $\beta$-adrenergic antagonists complement these effects by reducing vascular resistance and sympathetic drive~\citep{whelton20182017}. 

Multiple classes of drugs achieve BP reduction by inhibiting the renin–angiotensin–aldosterone system (RAAS), the central hormonal system regulating electrolyte balance~\citep{te2015hypertension}. Frequently used RAAS inhibitors include renin inhibitors, angiotensin-converting enzyme (ACE) inhibitors, angiotensin receptor blockers (ARB), and mineralocorticoid receptor antagonists~\cite[MRA,][]{carey2003intrarenal}. Within RAAS, aldosterone is a potent hormone that regulates sodium retention, acting primarily through the epithelial sodium channel (ENaC) in the distal nephron. By binding to mineralocorticoid receptors, aldosterone up-regulates ENaC expression and activity, and enhances sodium reabsorption and potassium secretion~\citep{lifton1996molecular}. Inhibition of ENaC, through amiloride and related potassium-sparing diuretics, blunts distal sodium uptake, enhances natriuresis, and lowers BP~\citep{canessa1994amiloride,tu2016triamterene}. Alternatively, blocking aldosterone signaling through MRA disrupts the activation of ENaC, reducing distal sodium reabsorption and thereby lowering BP, as shown in clinical trials~\citep{nishizaka2003efficacy,williams2015spironolactone}.

In individuals with higher aldosterone sensitivity, even a small increase in aldosterone can lead to significant BP elevation. This sensitivity was first discovered by studying the BP response to 9-$\alpha$ fludrocortisone (florinef $0.2$ mg daily), a powerful synthetic mineralocorticoid, at a dose doubling the endogenous aldosterone production~\citep{tu2014racial};  the biological processes were later delineated~\citep{gray2021aldosterone}. The fact that aldosterone sensitivity can vary by age and ethnicity~\citep{tu2014racial,nanba2017age,tu2018age} raises the possibility of heterogeneous treatment and/or mediation effects.

In summary, although the molecular targets and biological mechanisms of anti-HTN agents differ, their therapeutic effects inevitably converge on restoring sodium balance through renal pathways. Understanding which biological intermediates transmit the therapeutic benefits, therefore, becomes essential for optimizing individualized treatment. For this reason, individuals with impaired renal function -- whose sodium-handling capacity is already compromised -- represent a particularly informative population to investigate mediating pathways of treatment response. This is the approach we take in this research.

\subsection{Albuminuria reduction as potential mediator?}

HTN is a treatable condition. Even in patients with renal impairment, defined as estimated glomerular filtration rate (eGFR) $\le 60$ mL/min/$1.73$ m$^2$, BP can be properly managed. For these patients,  one hypothesized mediation mechanism involves the reduction of albuminuria. 

Albumin is a protein produced by the liver; its main function is to maintain BP and fluid volume. Healthy kidneys filter a very small amount of albumin. When the kidney is injured, abnormal amounts can leak into the urine, i.e., albuminuria. In clinical medicine, albuminuria is measured by the urine albumin-to-creatinine ratio~(UACR). Higher UACR indicates glomerular injury and endothelial dysfunction and is strongly prognostic for cardiovascular and renal events~\citep{barzilay2024albuminuria}. Therefore, early reductions in UACR following the initiation of anti-HTN therapy would indicate repaired kidney injury, and UACR reduction could potentially serve the role of a treatment‐responsive intermediate rather than a passive biomarker~\citep{molitoris2022albumin}.

The notion of UACR reduction being a viable mediator for anti-HTN therapies is also supported by empirical evidence. In recent clinical trials, different classes of drugs have been shown to derive BP benefits through UACR change. In a pooled analysis of two large phase 3 clinical trials of finerenone, a newer class of MRA, 4-month UACR reduction was found to mediate substantial proportions of both kidney and cardiovascular injuries in patients with type 2 diabetes~\citep{agarwal2023impact}. In a mechanistic analysis of data from another trial of anti-HTN treatments in patients with advanced CKD, thiazide-like diuretic chlorthalidone was found to lower BP through reduction of UACR~\citep{agarwal2024mechanisms}. Together, these data point to the plausibility of UACR change acting as a mediator for BP benefit in broader populations with diverse treatments.

Whether UACR reduction can mediate the effects of a wider range of antihypertensive treatments, instead of specific classes of drugs, in patients with renal impairment is a clinically important yet unanswered question. While the stipulation that most HTN therapies take effect through some renal mechanisms is biologically sound, whether UACR reduction conveys a signal for therapeutic effect is untested. A related question is whether the mediation effects of UACR are homogeneous across individuals. Earlier studies have shown that the BP effects of HTN drugs vary significantly across individuals~\citep{li2021robust,sundstrom2023heterogeneity}, so it is conceivable that the mediation effects also depend on patient characteristics. In case of such a heterogeneous mediation effect, it is logical to investigate what type of patients stand to benefit from this effect pathway. To elucidate, we analyze data from a large multicenter clinical trial.

\subsection{SPRINT: An HTN treatment trial}

The Systolic Blood Pressure Intervention Trial (SPRINT) was a landmark study that enrolled $9,361$ adults with essential HTN but without diabetes~\citep{sprint2015randomized}. Participants were recruited at $102$ clinical sites across the United States. Enrolled participants were randomly assigned to an intensive treatment arm, which set the systolic BP target to $\le 120$ mmHg, versus a usual care arm, with a systolic BP target of $\le 140$ mmHg. The study was approved by the institutional review board at each participating study site. The study reported that the intervention arm had significantly lower BP, fewer cardiovascular events, and lower all-cause mortality. Findings of the trial have since prompted a revision of practice guidelines with lowered BP targets~\citep{whelton20182017}. 

Notably, the SPRINT intervention did not mandate a specific anti-HTN regimen; clinicians were free to use various drug combinations to achieve BP targets. Consequently, the physiological pathways underlying the observed benefit could indeed vary and, as a result, remain incompletely characterized. The SPRINT study, nonetheless, presented a unique opportunity to investigate the mediation role of UACR in patients with compromised kidney function. We focus on this subpopulation because of its biological relevance, where renal autoregulation is disrupted and UACR reduction can plausibly mediate BP benefit. 

We identify participants with baseline eGFR $\le 60$ mL/min/$1.73$ m$^2$ in the SPRINT trial as our target population. This results in a subset of $n=1,963$ participants, $1,002$ of whom received the SPRINT intervention and $961$ received control treatment. Participants' baseline characteristics are presented as supplementary material in Section~\ref{appendix:sec:sprint}. The SPRINT trial collected extensive information on the participants. We group these characteristics into broad categories, including demographic and lifestyle variables, clinical measurements, major comorbidities, and laboratory measures (Tables~\ref{appendix:table:sprint_demo}--\ref{appendix:table:sprint_lab}).

A preliminary causal mediation analysis (CMA) is conducted using the existing method~\citep{imai2010general}.  Consistent with a previous study of UACR~\citep{agarwal2023impact}, we define the mediator as the change in logarithmic UACR from baseline to $6$ months ($M$), which captures early renal responses to treatment. Systolic BP at $12$ months is used as the primary outcome ($Y$); pre-treatment characteristics are considered as covariates ($\bZ$); see full list in Section~\ref{appendix:sub:sprint_charact}. The large set of covariates provides a comprehensive view of the factors that can influence the mediation effects of UACR in classic CMA. Table~\ref{table:sprint_cma} shows that the estimated natural indirect effect (NIE) differs markedly depending on whether covariates are included.

\begin{table}
    \begin{center}
        \caption{\label{table:sprint_cma} Causal mediation analysis of the SPRINT data: Summary of the population-average mediation effects with and without covariate adjustment.}
        \begin{tabular}{l r c c r c}
        \hline
        & \multicolumn{2}{c}{{Adjusted for Covariates}} && \multicolumn{2}{c}{Unadjusted} \\
        \cline{2-3} \cline{5-6}
        \multicolumn{1}{c}{\multirow{-2}{*}{Effect}}  & \multicolumn{1}{c}{Estimate} & \multicolumn{1}{c}{95\% CI} 
        && \multicolumn{1}{c}{Estimate} & \multicolumn{1}{c}{95\% CI} \\
        \hline
        ACME 
        & $-0.015$ & $(-0.031,\; 0.000)$ 
        && $-0.002$ & $(-0.015,\; 0.010)$ \\[3pt]
        
        ADE 
        & $-0.823$ & $(-0.907,\; -0.750)$ 
        && $-0.848$ & $(-0.931,\; -0.764)$ \\[3pt]
        
        Total Effect 
        & $-0.838$ & $(-0.918,\; -0.770)$ 
        && $-0.850$ & $(-0.932,\; -0.771)$ \\[3pt]
        
        Proportion Mediated 
        & $0.018$ & $(0.003,\; 0.040)$ 
        && $0.002$ & $(-0.012, \; 0.016)$ \\
        \hline
        \end{tabular}
    \end{center}
\end{table}

As shown in Table~\ref{table:sprint_cma}, when covariates are adjusted, the estimated NIE of the SPRINT intervention through UACR is small but statistically significant ($\mathrm{NIE}=-0.015$; $95\%$ CI: $(-0.031, 0.000)$; $p = 0.022$), indicating that UACR change may account for a measurable, albeit modest, portion of the treatment effect on 12-month systolic BP. In contrast, the unadjusted analysis yields an indirect effect essentially indistinguishable from zero ($\mathrm{NIE}=-0.002$; $95\%$ CI: $(-0.015, 0.010)$; $p = 0.744$), despite similar estimates of the direct and total effects between the two analyses. 

This discrepancy suggests that the mediation signal may be driven by covariates that differ across individuals, rather than a uniform population-wide path effect. This sensitivity to adjustment implies that the mediation effect may vary across patient subgroups defined by their risk profiles. This observation underscores the limitations of population-averaged CMA and motivates the need for methods capable of capturing heterogeneous mediation effects, allowing for the possibility that specific biological pathways, such as UACR reduction, operate only in subsets of patients rather than the entire population.

The large number of covariates poses a separate challenge:  Incorporating all covariates without selection may dilute true signals, inflate variance, and obscure heterogeneity in mediation effects. Moreover, high-dimensional adjustment also complicates the estimation of interaction structures that are central to heterogeneous mediation analysis. These considerations lead us to consider imposing regularization to stabilize estimation, reduce noise, and isolate covariates that meaningfully modify mediator–outcome or treatment–mediator relations. This is an issue addressed in the subsequent methodological development.

\subsection{Methodological challenges}

CMA is essential for delineating mechanisms by which an intervention exerts its influence. Over the past decades, a well-developed framework has emerged for estimating NIE under explicit identification assumptions~\citep{imai2010general,vanderweele2015explanation}. This framework has made CMA accessible for clinical studies~\citep{lee2021guideline} to evaluate complex multifactorial conditions such as HTN and CKD~\citep{agarwal2024mechanisms}.

Existing CMA techniques have mostly focused on population-average mediation effects, implying that the indirect pathway operates uniformly across individuals. This implicit assumption, however, is rarely tenable in studies of diseases with diverse mechanisms or treatments, where treatment interacts with patient characteristics and mediation pathways~\citep{li2021robust,wang2021causal}. HTN is emblematic of this complexity: physiologic heterogeneity and diverse renal and vascular mechanisms make it unlikely that any single mediator captures a universal pathway for all therapeutic benefits. As illustrated in our preliminary SPRINT analyses, population-average mediation effects can appear or disappear depending on covariate adjustment, suggesting strong underlying heterogeneity.

Methodologically, characterizing heterogeneous mediation effects remains challenging. Several authors have begun to explore subgroup-specific and individualized mediation effects~\citep{wang2021causal,huan2024individualized,ting2025estimating}, but the literature remains limited. Challenges persist in practical implementation:
\begin{enumerate}
    \item How to model heterogeneous effects. Heterogeneous NIEs arise through treatment–covariate, mediator–covariate, and/or treatment–mediator interactions~\citep{vanderweele2009conceptual}. Standard parametric linear structural equation modeling (LSEM) approaches can accommodate such structures in principle~\citep{van2019exploratory}, but in practice, these models become fragile when the number of covariates is large. Interaction terms increase combinatorially, leading to unstable estimates and violations of the hierarchical principle. Existing CMA methods do not provide tools for estimating mediator-related interactions at scale.

    \item How to properly select moderating factors. Modern clinical studies often collect large numbers of variables. Even when the number of covariates does not exceed the number of data units, thus not meeting the technical definition of high-dimensionality, inclusion of all patient characteristics can dilute true signals, inflate variance, and obscure meaningful variation in mediation~\citep{huang2016hypothesis}. While regularization methods can help, their integration into CMA has been uneven. Regularization complicates causal decomposition because penalties may shrink coefficients in ways that violate mediation identifiability; methods such as pathway lasso address high-dimensional mediators~\citep{zhao2022pathway}, but do not directly address the issues of effect heterogeneity and interaction.

    \item Lack of theoretically grounded inference procedures for heterogeneous mediation effects in high-dimensional models. Post-selection inference in the context of mediation analysis is rarely available. Without a careful quantification of the uncertainty, investigators cannot distinguish real mechanistic heterogeneity from noise.
\end{enumerate}

Together, these challenges have revealed a methodological gap: Although clinical questions frequently demand identification of patients who benefit through specific mechanisms, existing CMA methods are not yet equipped to estimate and infer individualized or subgroup-specific mediation effects, especially in situations of a large number of covariates.


\section{Methods}
\label{s:method}

\subsection{ Causal definitions and assumptions}

In this research, we consider studies with a binary treatment, where $t=1$ represents intervention and $t=-1$ represents control, with $\mathcal{T}=\{1,-1\}$ as the support of treatment. Let $M(t)$ be the potential value of the mediator when the treatment is set to $t$, and $\mathcal{M}$ be its support. Let $Y(t,M(t))$ be the potential outcome under treatment $t$ when the mediator takes the value of $M(t)$, with $\mathcal{Y}$ being its support. We write the vector of pretreatment covariates as  $\bZ\in\mathbb{R}^{p}$, where the first element is $1$ for the intercept, with $\mathcal{Z}$ being its support. Throughout, we allow the treatment and mediator effects to vary with covariates to accommodate heterogeneous causal pathways.

\subsubsection{Heterogeneous treatment and mediation effects}
  
\sloppy
With homogeneity, the average treatment effect (ATE) of the intervention is defined as $\tau=\mathbb{E}\left\{ Y(1,M(1))-Y(-1,M(-1))\right\}$. When treatment effects differ across individuals with covariates $\bz$, the average treatment effect can be written in a conditional form as
\begin{equation}\label{eq:def_hte}
    \tau(\bz)=\mathbb{E}\left\{Y(1,M(1))-Y(-1,M(-1))\mid \bZ=\bz \right\}.
\end{equation}
This conditional estimand captures the treatment heterogeneity over $\bz$ and reflects the individual response to the treatment.

With a mediator, the treatment effect decomposes into conditional direct and indirect effects. For $t\in \{1,-1\}$, we write
\begin{eqnarray}
    \delta(t\mid \bz) &=& \mathbb{E}\left\{Y(t,M(1))-Y(t,M(-1))\mid \bZ=\bz \right\}, \\
    \xi(t\mid \bz) &=& \mathbb{E}\left\{Y(1,M(t))-Y(-1,M(t))\mid \bZ=\bz \right\}.
\end{eqnarray}
The quantity $\delta(t\mid\bz)$ represents how the outcome would change if the mediator were shifted from the value it would take under control to the value it would take under intervention, holding treatment fixed at $t$. This is the conditional average indirect effect (cAIE). Extending the definitions in existing mediation literature, we have the conditional natural indirect effect~\citep{pearl2001direct} or the conditional pure indirect effect for $\delta(-1\mid\bz)$ and the conditional total indirect effect for $\delta(1\mid\bz)$~\citep{robins1992identifiability}. The quantity $\xi(t\mid \bz)$ captures the remaining conditional direct effect (cADE) not operating through the mediator. These effects satisfy
\begin{equation} 
    \tau(\bz)=\delta(1\mid\bz)+\xi(-1\mid\bz)=\delta(-1\mid\bz)+\xi(1\mid\bz),
\end{equation}
where the heterogeneous treatment effect is decomposed into heterogeneous indirect and direct effects.

\subsubsection{Identifiability assumptions}

Identification follows established causal mediation theory, which has been discussed extensively~\cite[examples include][among others]{robins1992identifiability,pearl2001direct,imai2010identification}. Here, we extend these to explicitly conditioning on $\bz$. 

\begin{description}
    \item[Assumption A1] \textbf{(SUTVA)} Stable Unit Treatment Value Assumption. 
    The treatment assignment mechanism remains the same across units, and each unit's potential outcomes depend only on their own treatment assignment.
    
    \item[Assumption A2] \textbf{(Positivity)}
    Let $\Omega_{tm}$ be the collection of $(t,m)$, where $t\in\mathcal{T}$ and $m\in\mathcal{M}$. For any subset $\mathcal{B}\in\Omega_{tm}$ with a positive measure, $\mathbb{P}\left[ (Z,M)\in\mathcal{B}\mid\bZ\right]>0$.

    \item[Assumption A3] \textbf{(Consistency)} If a subject receives treatment $T=t$, then $M=M(t)$ and $Y=Y(t,M(t))$, where $M$ and $Y$ are observed data.

    \item[Assumption A4] \textbf{(Treatment ignorability)} For any $m\in\mathcal{M}$,
        \begin{equation}
            \left\{Y(1,m),Y(-1,m),M(1),M(-1)\right\}~\indep~ T\mid \bZ.
        \end{equation}

    \item[Assumption A5] \textbf{(Sequential ignorability)} For each $t$ and each mediator value $m$, $Y(t,m)~\indep~M(t)\mid T=t,\bZ.$ 
    
    \item[Assumption A6] \textbf{(LSEM)} The following parametric LSEMs are assumed,
    \begin{eqnarray}
        M(t) &=& \balpha_{0}^\top\bZ +\balpha_{1}^\top\bZ t+\varepsilon(t), \label{eq:model_m} \\
        Y(t,m) &=& \bgamma_{0}^\top\bZ+\bgamma_{1}^\top\bZ t+\beta_{0} m+\beta_{1}mt+\eta(t,m), \label{eq:model_y}
    \end{eqnarray}
    where $\balpha_{0},\balpha_{1},\bgamma_{0},\bgamma_{1}\in\mathbb{R}^{p}$ and $\beta_{0},\beta_{1}\in\mathbb{R}$ are model coefficients, $\varepsilon(t)$ and $\eta(t,m)$ are independent errors with mean zero. 
    
\end{description}

The SUTVA assumption~\citep{rubin1980randomization} assumes consistency of the treatment regime across units and there is no interference. In the SPRINT application,  the consistency component of the assumption is satisfied because treatment protocols were clearly defined and uniformly implemented, ensuring the observed mediator and outcome correspond to the potential outcomes under the assigned treatment. In addition, the mediator and outcome are biologically subject-specific with no interference across participants.
The positivity assumption ensures sufficient support overlap for all covariate values with positive density. The positivity and consistency assumptions are standard in causal mediation analysis. 
Assumption A4 assumes that, given the pretreatment covariates, there exists no unmeasured mediator-treatment or outcome-treatment confounding. In our application, this assumption holds as the treatment assignment is randomized. 
Assumption A5 assumes that under the single-world intervention graphs~\citep{richardson2013single}, the potential outcomes of $Y$ and $M$ are conditionally independent. In other words, there exists no unmeasured mediator-outcome confounding. This assumption under the single-world setting is a relaxation of the cross-world independence assumption~\citep{robins2010alternative} given the proposed parametric LSEMs~\cite[Assumption A6, see a discussion in][]{andrews2021insights}. Assumption A6 specifies a linear system where heterogeneity arises through interactions between treatment, mediator, and covariates. Specifically, $\bZ t$ in both models represents a key source of heterogeneity. The coefficient $\balpha_{1}$ quantifies how the treatment modifies the mediator differently across the covariate space. The coefficients $\bgamma_{1}$ and $\beta_{1}$ describe covariate-modified direct and mediator–treatment interactions influencing the outcome.

Under the above assumptions, the cAIE and cADE can be identified from the observed data and have closed-form expressions (Theorem~\ref{thm:causal_def}). The proof of the theorem is provided in Section~\ref{appendix:sub:proof_causal_def} of the supplementary materials. These provide individualized mediation contrasts that depend on the treatment assignment and the covariate profile $\bz$, making it possible to identify subgroups where the mediator is mechanistically active.

\begin{theorem}\label{thm:causal_def}
    Under Assumptions A1--A6, for $t\in\{1,-1\}$,
    \begin{eqnarray*}
        \tau(\bz) &=& \mathbb{E}\left\{Y(1,M(1))-Y(-1,M(-1))\mid \bZ=\bz \right\}=2\bgamma_{1}^\top\bz+2\beta_{0}\balpha_{1}^\top\bz+2\beta_{1}\balpha_{0}^\top\bz,\\
        \delta(t\mid\bz) &=& \mathbb{E}\left\{Y(t,M(1))-Y(t,M(-1))\mid \bZ=\bz \right\}= 2(\beta_{0}+\beta_{1}t)\balpha_{1}^\top\bz, \\
        \xi(t\mid\bz) &=& \mathbb{E}\left\{Y(1,M(t))-Y(-1,M(t))\mid \bZ=\bz \right\} = 2\bgamma_{1}^\top\bz+2\beta_{1}(\balpha_{0}^\top\bz+\balpha_{1}^\top\bz t).
     \end{eqnarray*}
\end{theorem}

\subsection{Model parameter estimation}
\label{sub:method}

This section introduces how to estimate the parameters in the proposed LSEMs and how to handle the interactions efficiently in the presence of a large number of covariates. A central challenge is that both models contain interactions between covariates and treatment, and the outcome model additionally includes an interaction between the mediator and treatment. In conventional parametric regression, the inclusion of interaction terms enforces a hierarchical principle: Whenever an interaction is included, the associated main effects must also be present, regardless of its significance~\citep{bien2013lasso}. With a large number of covariates, this hierarchy makes model fitting unstable and complicates the use of penalization methods.

\subsubsection{A modified-covariate approach}

To avoid imposing such a hierarchical constraint and to simplify the interaction structures, we adopt a modified covariate strategy proposed by \citet{tian2014simple}. The central idea is to recast the original model into separate linear regressions within the intervention and control groups. This works for binary treatment assignment $t\in \{-1,1\}$ as the interaction terms, $\bZ t$ and $mt$, can be represented through group-specific regression coefficients.

Since models~\eqref{eq:model_m} and~\eqref{eq:model_y} face the same challenge, for narrative convenience, we present the estimation strategy in a generic linear regression model, which applies to both
\eqref{eq:model_m} and~\eqref{eq:model_y}
\begin{equation}\label{eq:model_general}
    R = \bphi_{0}^\top\bS + \bphi_{1}^\top\bS T +\varepsilon,
\end{equation}
where $R$ is a continuous scalar outcome (either $M$ or $Y$), $\bS\in\mathbb{R}^{q}$ is a covariate vector with first element one, $\bS=\bZ$ ($q=p$) for the mediator model and $\bS=(\bZ^\top,M)^\top$ ($q=p+1$) for the outcome model, and $\varepsilon$ is an error term with mean zero. 
With this formulation, for a balanced design, $\bphi_{0}$ captures the average outcome over both groups and $\bphi_{1}$ captures the treatment effect. 
For a study of $n$ subjects, let $\{T_{i},R_{i},\bS_{i}\}$ be the observed data from unit $i$ and $\varepsilon_{i}$ as the model error, for $i=1,\dots,n$. Partition the data into two groups: the intervention group ($T=1$) with $n_{1}$ subjects and the control group ($T=-1$) with $n_{-1}$ subjects. Let $\bR_{(1)}\in\mathbb{R}^{n_{1}}$ be the vector of outcomes for those units in the intervention group, $\bS_{(1)}\in\mathbb{R}^{n_{1}\times q}$ be the matrix of covariates, and $\bvarepsilon_{(1)}\in\mathbb{R}^{n_{1}}$ be the vector of model errors, and define $\bR_{(-1)},\bS_{(-1)},\bvarepsilon_{(-1)}$ for the control group analogously. Model~(\ref{eq:model_general}) can be rewritten as
\begin{eqnarray}
    \bR_{(1)} = \bS_{(1)}\bphi_{(1)}+\bvarepsilon_{(1)} &\text{ with }& \bphi_{(1)}=\bphi_{0}+\bphi_{1}, \label{eq:model_case} \\
    \bR_{(-1)} = \bS_{(-1)}\bphi_{(-1)}+\bvarepsilon_{(-1)} &\text{ with }& \bphi_{(-1)}=\bphi_{0}-\bphi_{1}, \label{eq:model_ctrl}
\end{eqnarray}
where the interaction term is absorbed into group-specific slopes.
This reparameterization yields a regression problem free of explicit interaction terms, eliminating the need to impose hierarchical constraints, facilitating the use of regularization approaches when the covariate dimension is high. 
Instead, two linear models will be fitted, one for each arm. The following relations recover the main and interaction coefficients, 
\begin{equation}\label{eq:phi_relation}
    \bphi_{0}=\frac{1}{2}\left(\bphi_{(1)}+\bphi_{(-1)} \right), \quad \bphi_{1}=\frac{1}{2}\left(\bphi_{(1)}-\bphi_{(-1)} \right).
\end{equation}

For estimation, we stack the treated and control samples:
\[
    \tilde{\bR}=\begin{pmatrix}
        \bR_{(1)} \\
        \bR_{(-1)}
    \end{pmatrix}\in\mathbb{R}^{n}, ~ \tilde{\bS}=\begin{pmatrix}
         \bS_{(1)} & \bzero \\
         \bzero & \bS_{(-1)}
    \end{pmatrix}\in\mathbb{R}^{n\times 2q}, ~ \bphi=\begin{pmatrix}
         \bphi_{(1)} \\
         \bphi_{(-1)}
    \end{pmatrix}\in\mathbb{R}^{2q}, ~ \tilde{\bvarepsilon}=\begin{pmatrix}
         \bvarepsilon_{(1)} \\
         \bvarepsilon_{(-1)}
    \end{pmatrix}\in\mathbb{R}^{n}.
\]
Models~\eqref{eq:model_case} and~\eqref{eq:model_ctrl} together become
\begin{equation}\label{eq:model_modified}
    \tilde{\bR} = \tilde{\bS}\bphi+\tilde{\bvarepsilon},
\end{equation}
which has no interaction terms. When the number of predictors is moderate and $n>2q$, ordinary least squares (OLS) can be used to estimate $\bphi$. 
\begin{equation}\label{eq:phi_OLS}
    \hat{\bphi}^{\text{OLS}}=\underset{\bphi}{\arg\min}\frac{1}{2}\|\tilde{\bR}-\tilde{\bS}\bphi\|_{2}^{2}.
\end{equation}
$\bphi_{0}$ and $\bphi_{1}$ can be then ascertained from~\eqref{eq:phi_relation}, and consequently, $(\balpha_0, \balpha_1, \bgamma_0, \bgamma_1, \beta_0,\beta_1)$ in the original models can be recovered.
However, under the high-dimensional setting, where the number of covariates is relatively large compared to the sample size ($2q\gg n$), OLS becomes less stable or even infeasible. We therefore introduce a regularization strategy.

\subsubsection{Generalized lasso for high-dimensional estimation}


To impose regularization, the target coefficients are those in the original models~\eqref{eq:model_m}--\eqref{eq:model_y}. With the relationships to the coefficients in the reparameterized model, the generalized lasso~\citep{tibshirani2011solution} is employed.
\begin{equation}\label{eq:phi_genlasso}
    \underset{\bphi}{\text{minimize}} ~\frac{1}{2}\|\tilde{\bR}-\tilde{\bS}\bphi\|_{2}^{2}+\lambda\|\bD\bphi\|_{1},
\end{equation}
where $\lambda\geq 0$ is the tuning parameter and $\bD\in\mathbb{R}^{r\times 2q}$ is the regularization matrix. 
The matrix $\bD$ is constructed to impose sparsity on $\bphi_0$ and $\bphi_1$ to select a parsimonious set of covariates using the relation in~\eqref{eq:phi_relation}. Specifically, we set
\begin{equation}\label{eq:Dmatrix}
    \bD=\begin{pmatrix}
         \bzero_{q-1} & \matI_{q-1} & \bzero_{q-1} & \matI_{q-1} \\
         \bzero_{q-1} & \matI_{q-1} & \bzero_{q-1} & -\matI_{q-1}
    \end{pmatrix}\in\mathbb{R}^{2(q-1)\times 2q}.
\end{equation}
When $\lambda=0$, the generalized lasso estimator is equivalent to the OLS estimator. When $\lambda>0$, the optimization yields a sparse solution.
The tuning parameter, $\lambda$, can be chosen based on the Bayesian information criterion (BIC), generalized cross-validation (GCV), or Mallow's Cp. In our numerical studies, we use Mallow's Cp as recommended in \citet{arnold2016efficient}, which performs well empirically and offers computational simplicity.


\subsubsection{Mediation parameters and cAIE estimation}

With $\hat{\bphi}_{(1)}$ and $\hat{\bphi}_{(-1)}$, from either the OLS or the generalized lasso, we can recover 
$$
\hat{\bphi}_{0}=\frac{1}{2}\left(\hat{\bphi}_{(1)} +\hat{\bphi}_{(-1)}\right) \textrm{  and  } \hat{\bphi}_{1}=\frac{1}{2}\left(\hat{\bphi}_{(1)} -\hat{\bphi}_{(-1)}\right).
$$
These yield direct estimates of $\hat{\balpha}_0$ and $\hat{\balpha}_1$ from the mediator model and $\hat{\bgamma}_0$, $\hat{\bgamma}_1$, $\hat{\beta}_0$, $\hat{\beta}_1$ from the outcome model. With these estimates, the individualized effects $\hat{\delta} (t\mid \bz)$ and $\hat{\xi} (t\mid \bz)$ follow from the closed-form expressions in Theorem~\ref{thm:causal_def}.

\subsection{Asymptotic properties}
\label{sub:asmp}

For the large-sample behavior of the proposed estimators, because the mediator and outcome models can be estimated either by OLS or by generalized lasso, depending on dimensionality, we present the two cases separately. The derivations for the results appear in Sections~\ref{appendix:sub:proof_OLS_asmp} and~\ref{appendix:sub:genlasso_asmp} of the supplementary materials.

\subsubsection{Asymptotics for OLS estimator}

Assuming $\Var(\varepsilon)=\sigma_{m}^{2}<\infty$ and $\Var(\eta)=\sigma_{y}^{2}<\infty$. Under the sequential ignorability assumption, the two errors are independent. Let

\[
   \bTheta=\begin{pmatrix}
       \balpha_{0} & \bgamma_{0} \\
       \balpha_{1} & \bgamma_{1} \\
       0 & \beta_{0} \\
       0 & \beta_{1}
   \end{pmatrix}, \quad \bSigma=\begin{pmatrix}
        \sigma_{m}^{2} & 0 \\
        0 & \sigma_{y}^{2}
   \end{pmatrix}.
\]
The theorem below gives the asymptotic distribution of the OLS estimator and that of the corresponding cAIE and cADE estimator with treatment assignment $t$ and covariates $\bz$.
\begin{theorem}\label{thm:OLS_asmp}
    For fixed $p<n$, assume $\bZ^\top\bZ/n\rightarrow\bQ$ as $n\rightarrow \infty$, and the distribution of the treatment assignment regime has a finite fourth-order moment. Then, the OLS estimator of $\bTheta$ converges asymptotically to a normal distribution.
    \begin{equation}
        \sqrt{n}~\mathrm{vec}(\hat{\bTheta}-\bTheta)\overset{\mathcal{D}}{\longrightarrow}\mathcal{N}\left(\bzero,\bSigma\otimes\bQ_{X}^{-1} \right),
    \end{equation}
    where $\mathrm{vec}()$ is the vectorization of a matrix, $\otimes$ is the Kronecker product operator, and $\bQ_{X}$ is provided in Section~\ref{appendix:sub:proof_OLS_asmp} of the supplementary materials. With a given treatment assignment $t$ and a vector of covariates $\bz$, the OLS-based estimator of cAIE and cADE has asymptotic normal distributions.
    \begin{equation}
        \sqrt{n}(\hat{\delta}-\delta)\mid t,\bz \overset{\mathcal{D}}{\longrightarrow} \mathcal{N}\left(0,\bH_{1}^\top(\bSigma\otimes\bQ_{X}^{-1})\bH_{1}\right), ~ \sqrt{n}(\hat{\xi}-\xi)\mid t,\bz \overset{\mathcal{D}}{\longrightarrow} \mathcal{N}\left(0,\bH_{2}^\top(\bSigma\otimes\bQ_{X}^{-1})\bH_{2}\right),
    \end{equation}
    where $\bH_{1}$ and $\bH_{2}$ are given in Section~\ref{appendix:sub:proof_OLS_asmp} of the supplementary materials.
\end{theorem}
These results imply that, in low-dimensional cases (where OLS is used), the mediation parameters cAIE and cADE, can be estimated at the parametric $n^{-1/2}$ rate with a tractable covariance structure. Standard Wald confidence intervals then follow directly. The expressions show that the uncertainty depends on both the mediator and outcome model fits, and on the covariate vector $\bz$ at which the heterogeneous effect is evaluated.

\subsubsection{Asymptotics for generalized lasso estimator}

With the added regularization, the asymptotics of the generalized lasso estimator depends on regularity conditions analogous to those described by \citet{lee2015model}, see Section~\ref{appendix:sub:genlasso_asmp} of the supplementary materials. Briefly, a restricted strong convexity (RSC) condition on the loss function and an irrepresentability-type condition with respect to $\bD$ are needed.

Under these conditions, the generalized lasso estimator yields both estimation and model selection consistency~\citep{lee2015model}.

\begin{theorem}\label{thm:genlasso_asmp}
    Under the regularity conditions in Section~\ref{appendix:sub:genlasso_asmp} of the supplementary materials, for some $0<\kappa_{1},\kappa_{2},\kappa_{3}<\infty$ and $\lambda=(8\kappa_{1}\sigma/\iota)\sqrt{\log(2q)/n}$, the estimator of~\eqref{eq:phi_genlasso} is unique, where $\sigma$ is the standard deviation of model error $\tilde{\varepsilon}$ and $\iota$ is defined in Section~\ref{appendix:sub:genlasso_asmp}. In addition, with probability at least $1-2(2q)^{-1}$, (1) the estimator is consistent:
    \[
        \|\hat{\bphi}-\bphi^{*}\|_{2}\leq \frac{4}{m}(\kappa_{3}+4\kappa_{1}\kappa_{2}/\iota)\sigma\sqrt{\frac{\log(2q)}{n}},
    \]
    and (2) model selection is consistent: $\hat{\bphi}\in\mathscr{M}$, where $m$, and $\mathscr{M}$ are defined in Section~\ref{appendix:sub:genlasso_asmp}.
    With a given treatment assignment $t$ and a vector of covariates $\bz$, the generalized lasso-based estimator of cAIE and cADE is consistent.
\end{theorem}
The proof and values of $\kappa_{1},\kappa_{2},\kappa_{3}$ are provided in Section~\ref{appendix:sub:genlasso_asmp} of the supplementary materials.
The convergence rate of the estimator is standard in high-dimensional regression and reflects the price of estimating models with many covariates and interactions. Importantly, the result ensures that heterogeneous mediation effects, which are computed from a combination of regression coefficients, are also consistent. This guarantees that the method does not invent spurious heterogeneity as the sample size grows. Unlike the OLS case, the generalized lasso estimator does not admit a closed-form asymptotic distribution for Wald intervals. For inference, we therefore rely on the multiple sample-splitting procedure described in the next section.


\subsection{Inference}
\label{sub:inference}

For OLS estimators, valid inference follows from Theorem~\ref{thm:OLS_asmp}. In this section, we focus on the inference for the generalized lasso estimators, which relies on a multiple splitting procedure. With a high probability, the generalized lasso screens in all relevant predictors under the regularity conditions. However, the regularized coefficients themselves are biased and not suitable for inference.
One way to reduce the bias is to refit the regression model. The strategy of multiple sample splitting has been utilized to avoid the over-optimism of using the data twice for both selection and post-selection inference~\citep{dezeure2015high,dai2023false}. 
With the generalized lasso regularization, in case the number of selected features still exceeds the sample size, a small ridge penalty is typically used to ensure estimation stability. Although this may introduce bias, as the sample size increases, it converges to the truth. 
The procedure for generalized lasso inference is summarized in Algorithm~\ref{alg:splitting}.
In Section~\ref{sec:sim}, the numerical performance of this procedure is evaluated through simulation.

\begin{algorithm}[t]
\caption{\label{alg:splitting}Multiple Sample-Splitting Inference for Generalized Lasso Estimators}

    \begin{algorithmic}[1]
    \item[Step 1.] For the $b$th iteration, randomly split the dataset, $\mathcal{D}$, into two subsets of equal size, $\mathcal{D}_{1}^{(b)}$ and $\mathcal{D}_{2}^{(b)}$. 

       \item[Step 2.] Use $\mathcal{D}_{1}^{(b)}$ to fit the mediator and outcome models via the generalized lasso, obtaining a set of selected covariates. Then refit the models on $\mathcal{D}_{2}^{(b)}$ using the selected variables with a small ridge penalty to ensure numerical stability.
       
        \item[Step 3.] Use refitted coefficients, compute the cAIE ($\hat{\delta}_i^{(b)}$) and cADE ($\hat{\xi}_i^{(b)}$) for eath individual $i\in\{1,\dots,n\}$ according to the closed-form expressions in Theorem~\ref{thm:causal_def}.
    
        \item[Step 4.] Repeat Steps 1--3 for $B$ times. This yields $B$ sets of effect estimates for each individual.
        
        \item[Step 5.] Combine the $B$ estimates to form point estimates and confidence intervals.  
    \end{algorithmic}
\end{algorithm}



\section{Simulation Studies}
\label{sec:sim}

To evaluate the performance of the proposed methods, we simulate data resembling the SPRINT trial in dimensionality and sparsity. We consider $100$ baseline covariates, $50$ of which follow a standard normal $\mathcal{N}(0,1)$ distribution, and $50$ follow a Bernoulli($0.5$) distribution. Treatment $T\in\{1,-1\}$ is generated randomly from Bernoulli($0.5$). With $100$ covariates, the actual number of predictors is $2p=200$, including the interaction terms. The mediator ($M$) and outcome ($Y$) are generated from models~\eqref{eq:model_m} and~\eqref{eq:model_y}; $95\%$ of the coefficients $\balpha_{0}$, $\balpha_{1}$, $\bgamma_{0}$, and $\bgamma_{1}$ are set to $0$ for sparsity. Model errors $\varepsilon$ and $\eta$ are independently generated from $\mathcal{N}(0, 0.5^{2})$. This setting induces treatment–covariate and mediator–covariate interactions, making population-average CMA insufficient while allowing direct evaluation of cADE/cAIE recovery. 

We compare the OLS and GenLasso (generalized lasso) across sample sizes $n=200$, $1000$, and $2000$. OLS is feasible only when $n>2p$; GenLasso is evaluated in all cases with tuning via Mallow's Cp~\citep{arnold2016efficient}. Inference for GenLasso uses the sample-splitting procedure introduced in Section~\ref{sub:inference} with $B=500$.

\begin{figure}
    \begin{center}
        \subfloat[cAIE: OLS]{\includegraphics[width=0.25\textwidth]{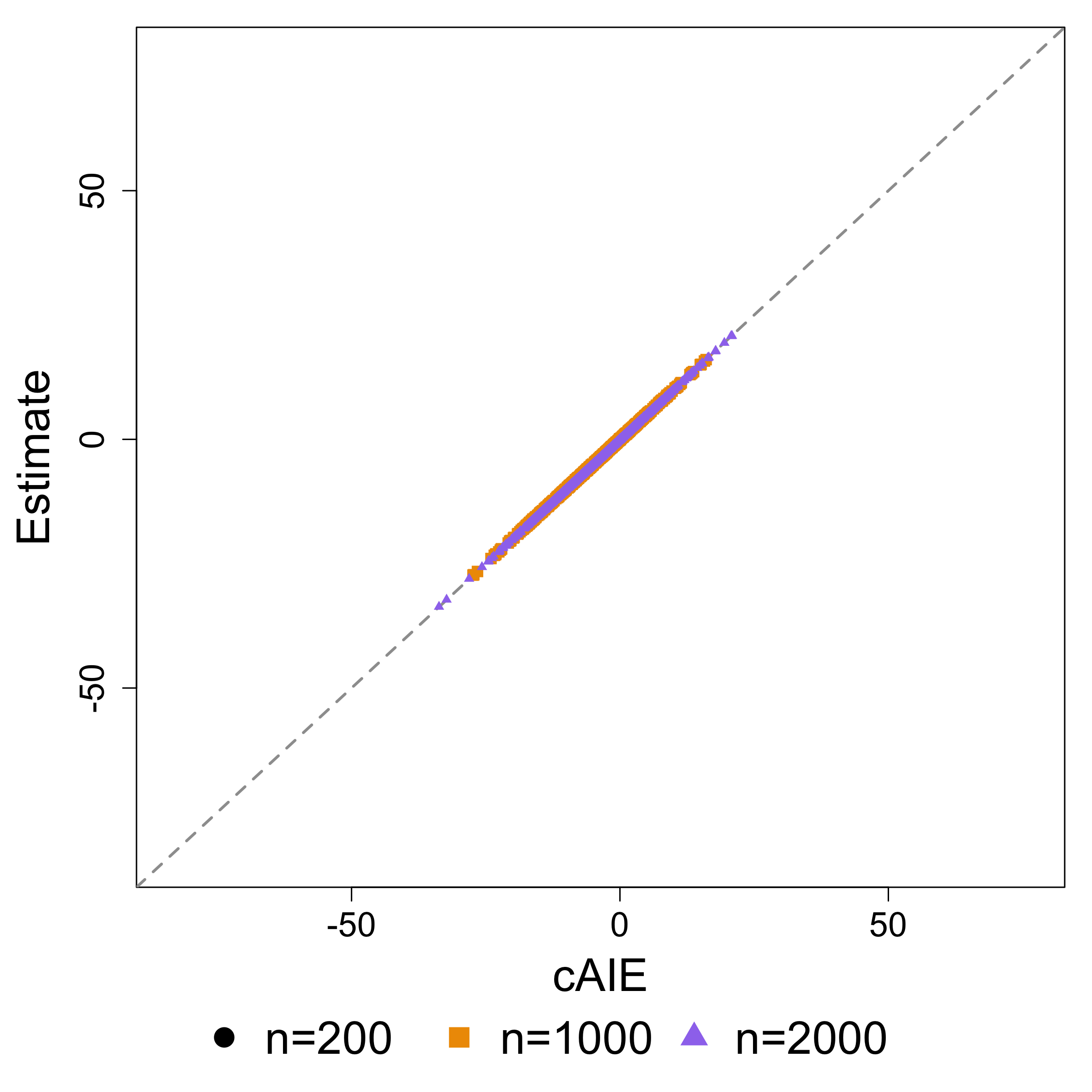}}
        \subfloat[cAIE: GenLasso]{\includegraphics[width=0.25\textwidth]{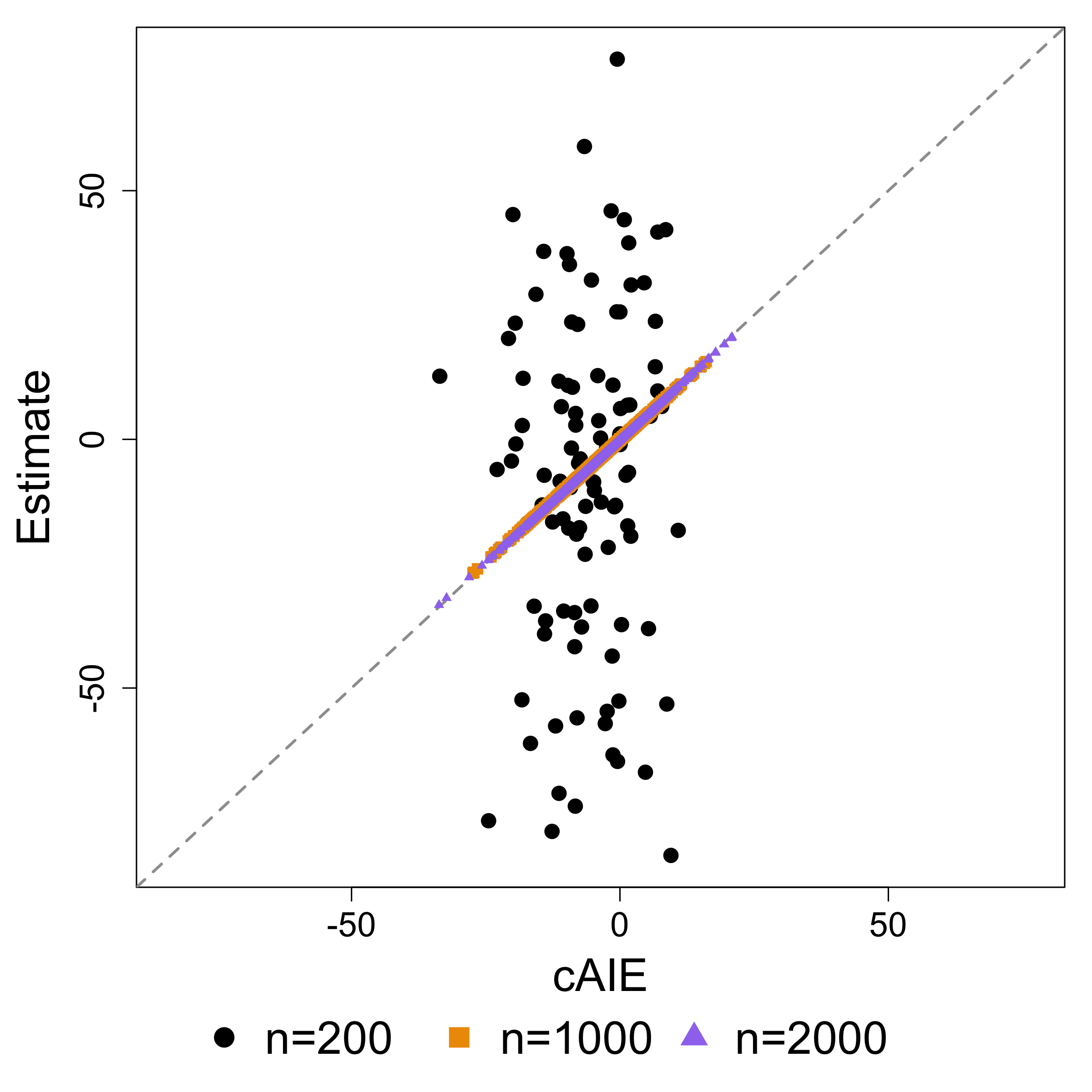}}
        \subfloat[cAIE: Bias]{\includegraphics[width=0.25\textwidth]{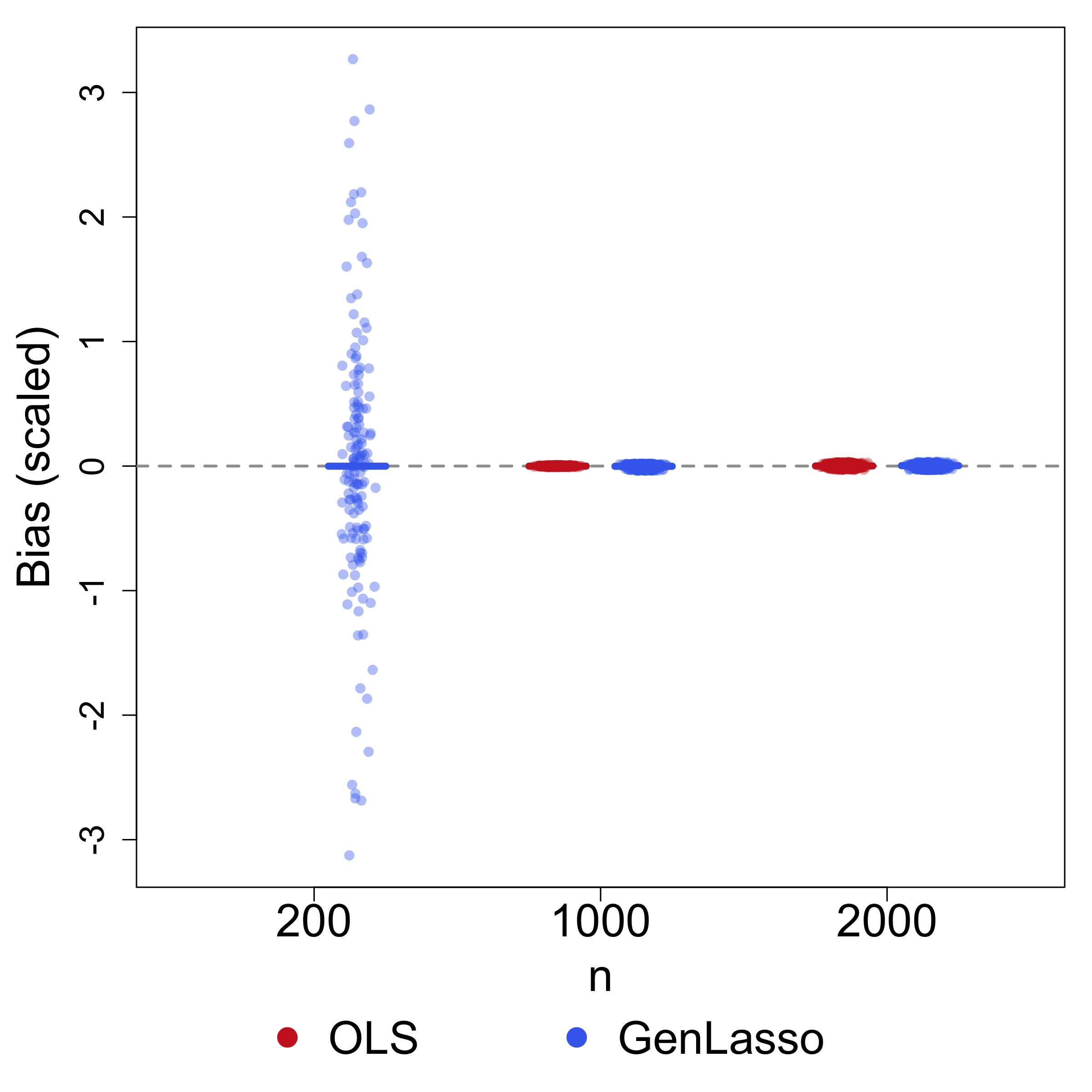}}
        \subfloat[cAIE: MSE]{\includegraphics[width=0.25\textwidth]{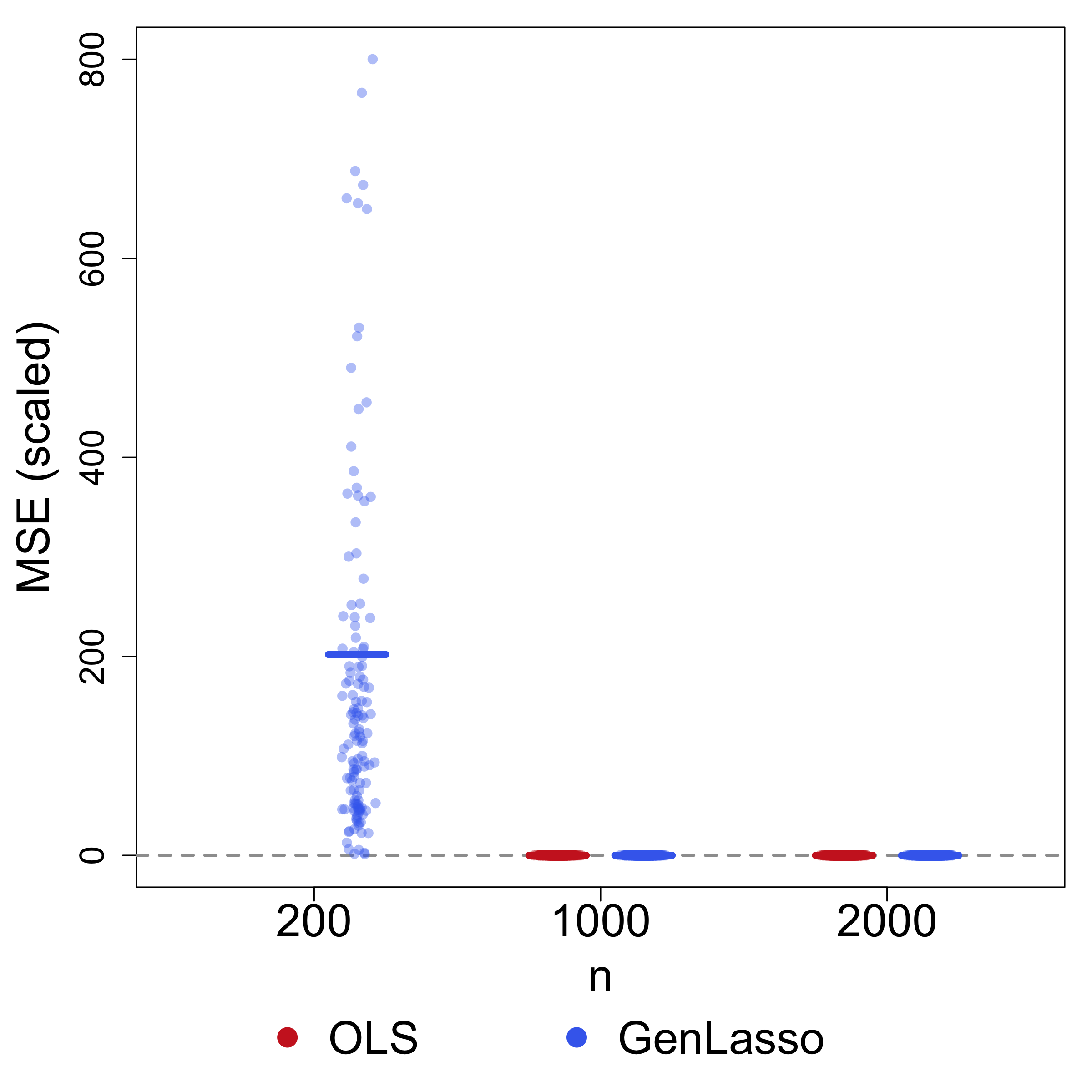}}

        \subfloat[cADE: OLS]{\includegraphics[width=0.25\textwidth]{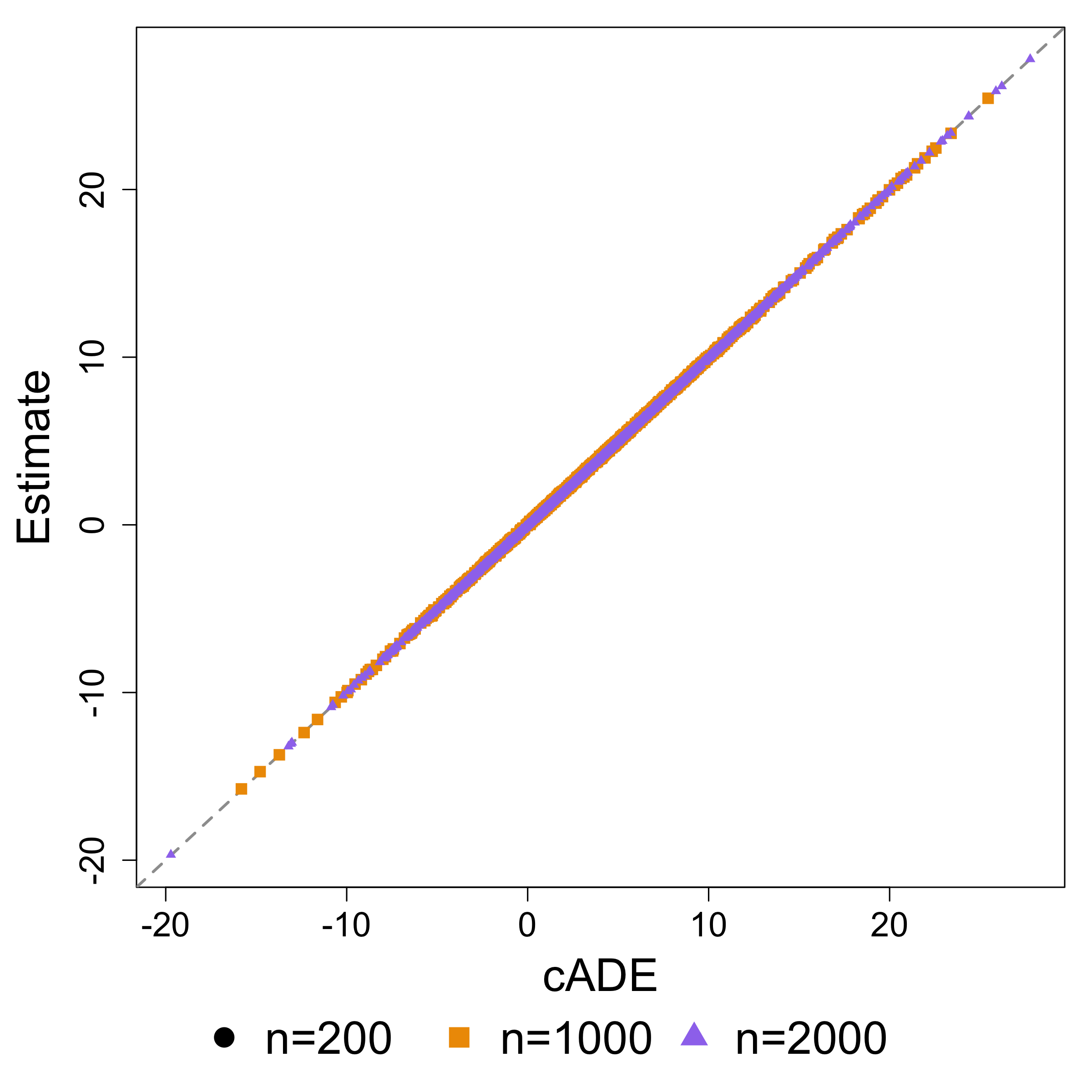}}
        \subfloat[cADE: GenLasso]{\includegraphics[width=0.25\textwidth]{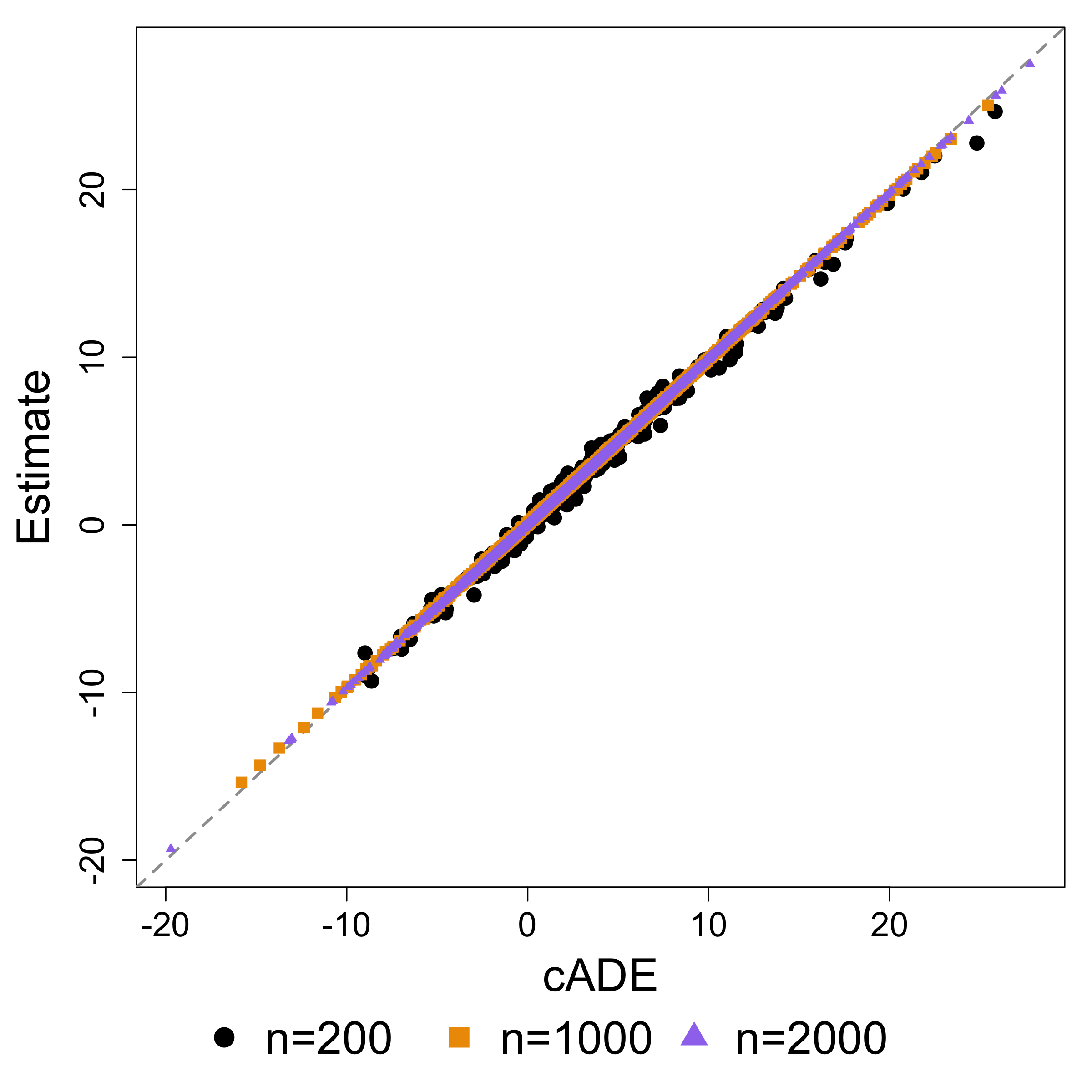}}
        \subfloat[cADE: Bias]{\includegraphics[width=0.25\textwidth]{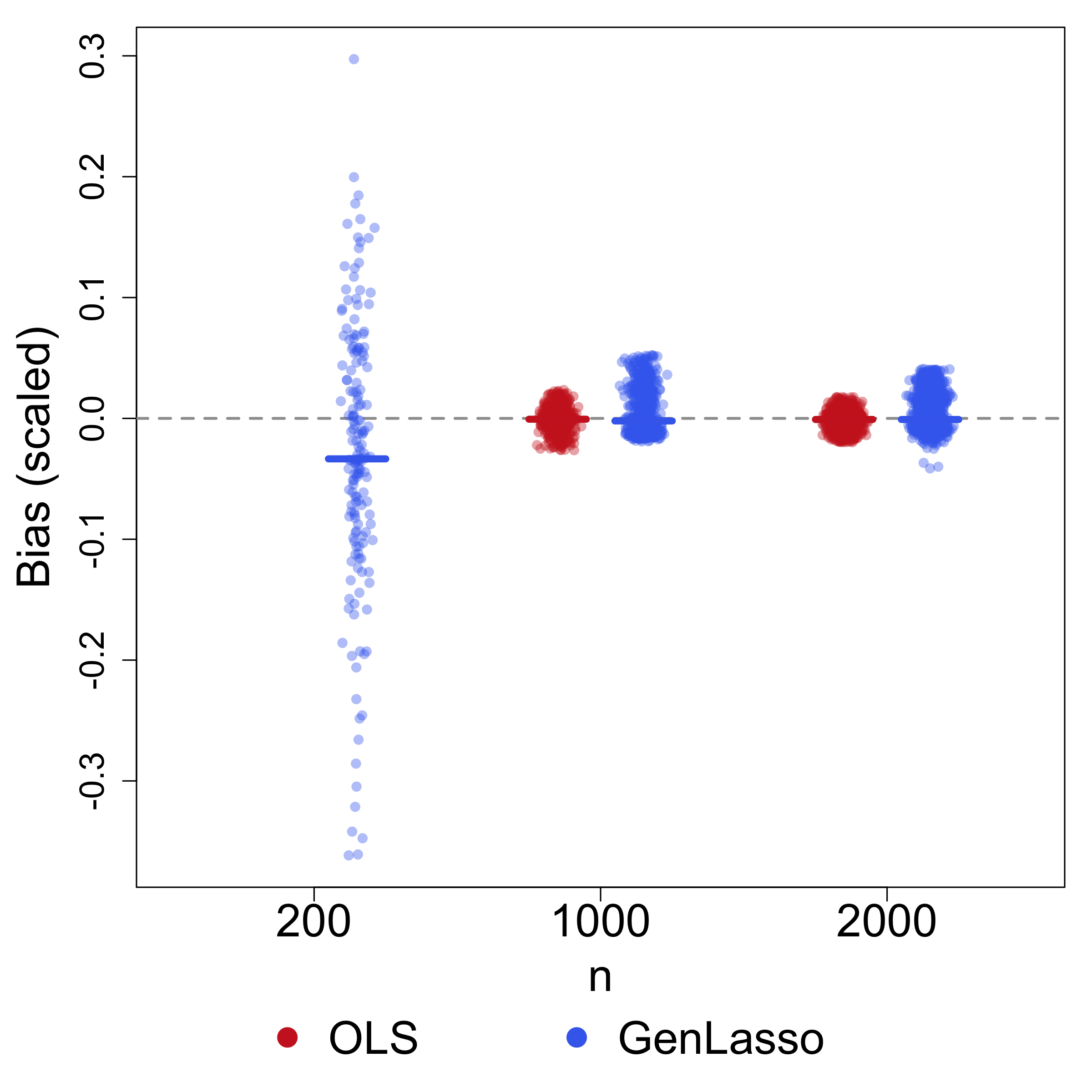}}
        \subfloat[cADE: MSE]{\includegraphics[width=0.25\textwidth]{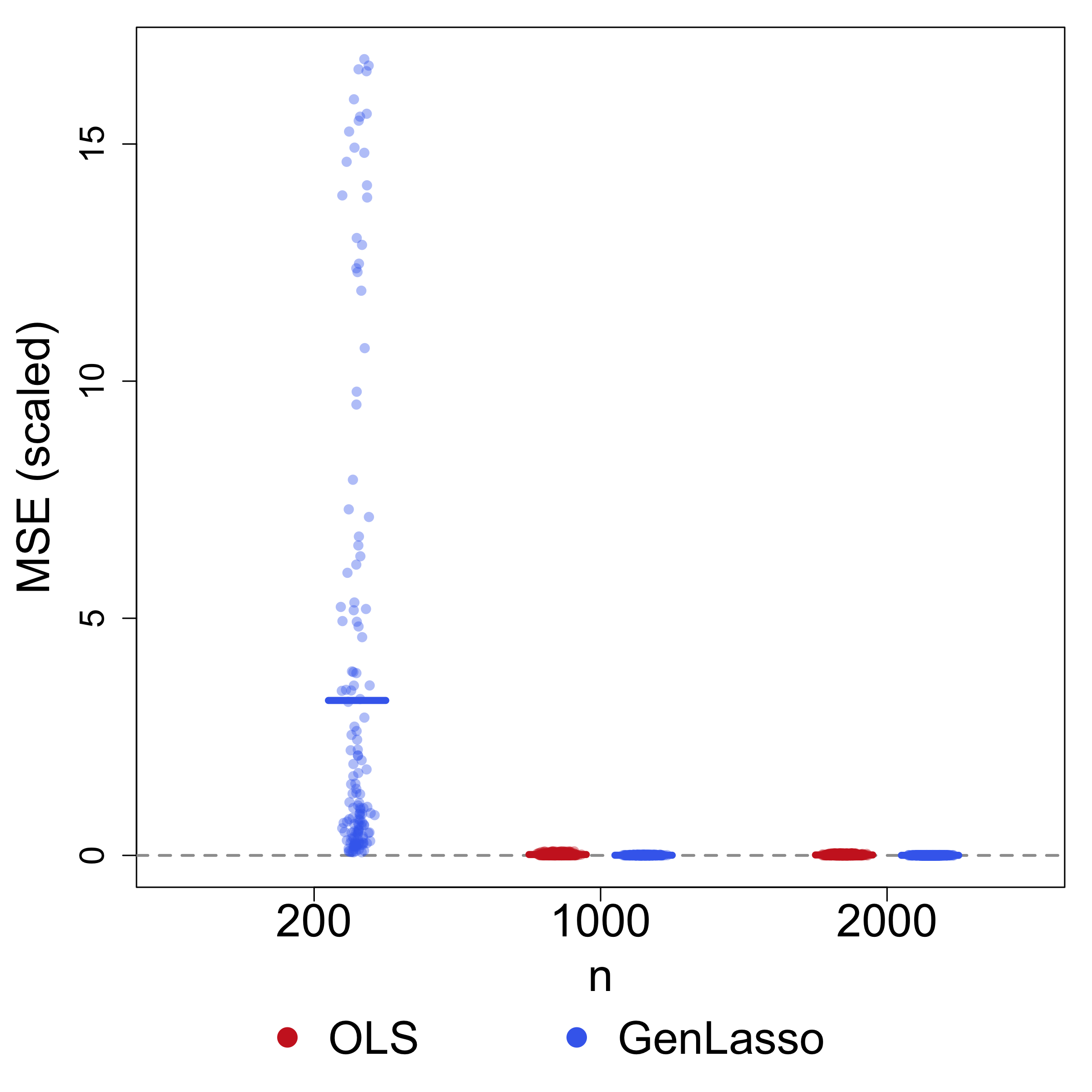}}
    \end{center}
    \caption{\label{fig:sim_est} \small{Simulation studies: The estimate and scaled bias and mean squared error (MSE) of cAIE and cADE from the OLS and GenLasso approaches across sample sizes $n=200,1000,200$ with $p=100$. Results are from $200$ simulated replications. In (c)--(d) and (g)--(h), the horizontal bar indicates the median values.}}
\end{figure}

Figure~\ref{fig:sim_est} shows that OLS (when feasible) and GenLasso recover heterogeneous mediation effects accurately. When $n=1000$ or $2000$, both methods produce estimates closely resembling the true cAIE and cADE, with low bias and small mean squared error (MSE). Under the high-dimensional condition ($n=200$), OLS is no longer feasible, while GenLasso performs reasonably well, with biases decreasing steadily as sample size increases.

\begin{figure}
    \begin{center}
        \subfloat[cAIE: CI size]{\includegraphics[width=0.25\textwidth]{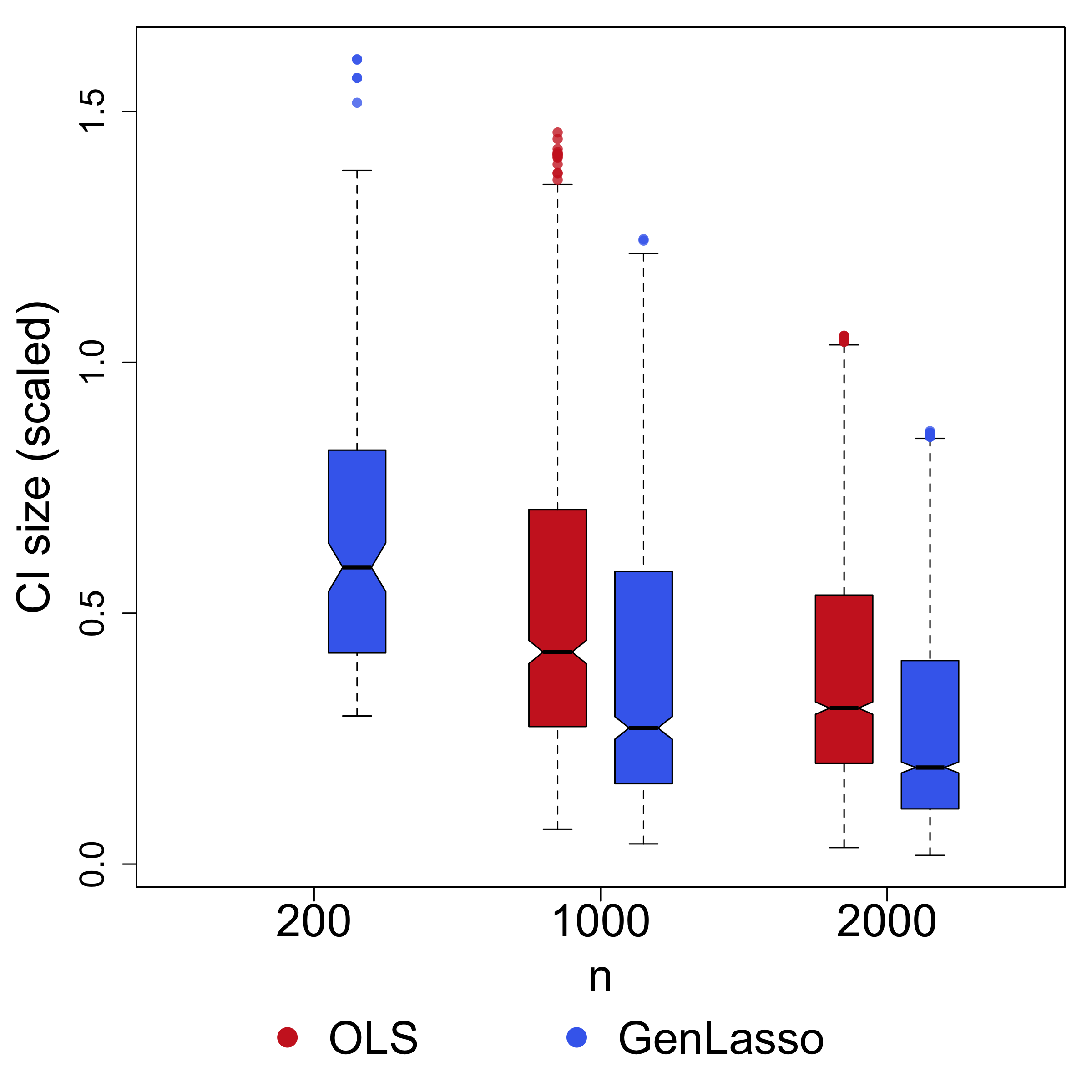}}
        \subfloat[cAIE: CP]{\includegraphics[width=0.25\textwidth]{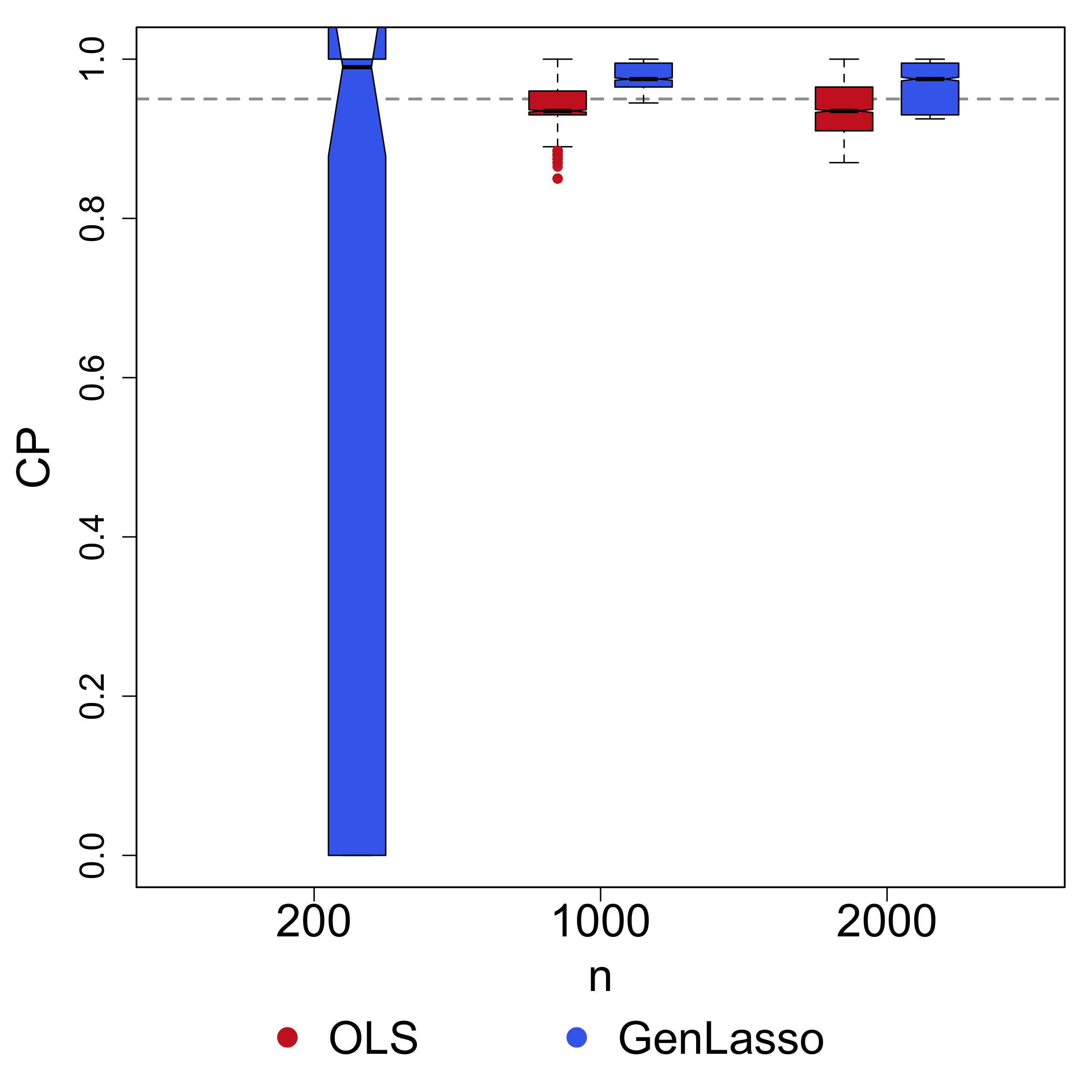}}
        \subfloat[cADE: CI size]{\includegraphics[width=0.25\textwidth]{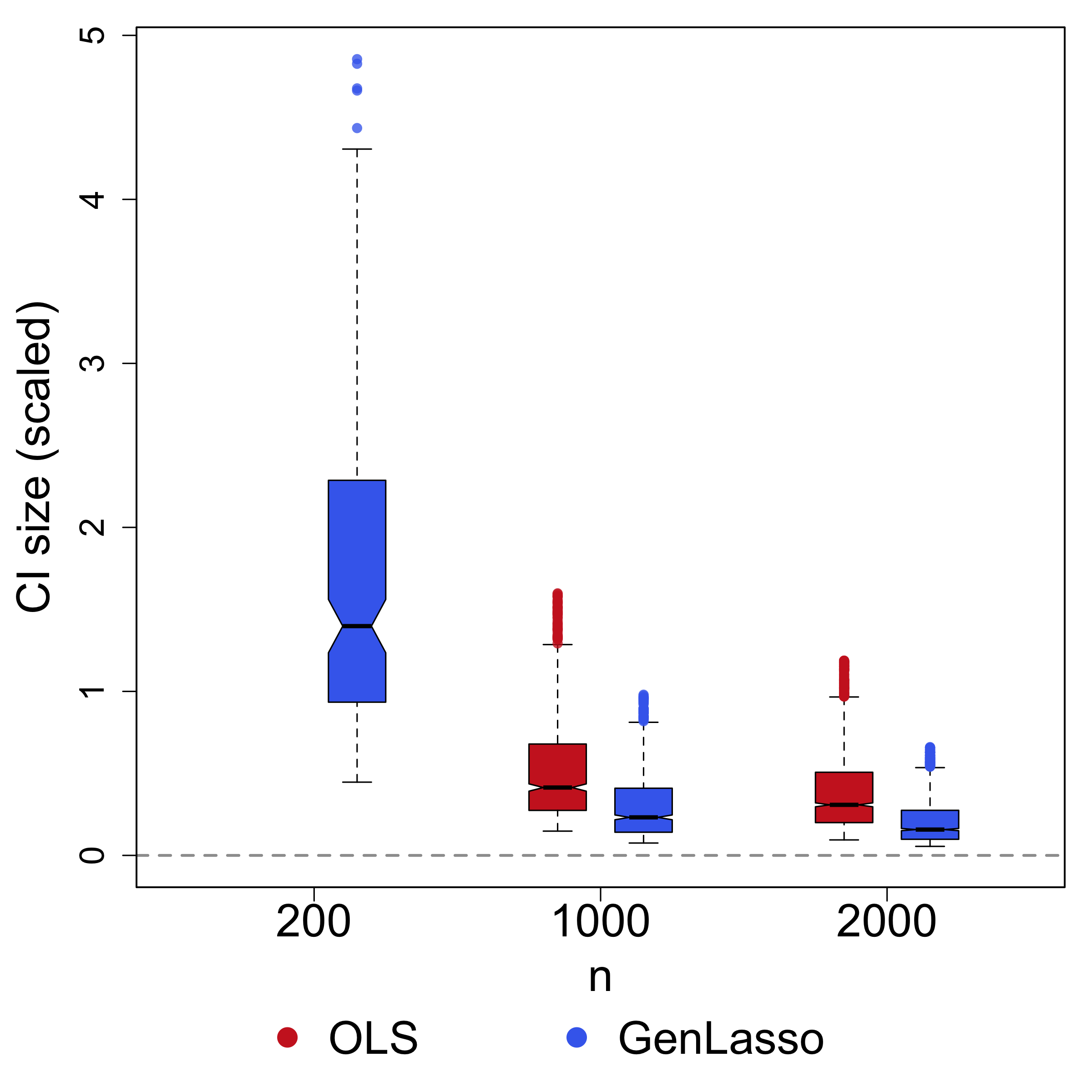}}
        \subfloat[cADE: CP]{\includegraphics[width=0.25\textwidth]{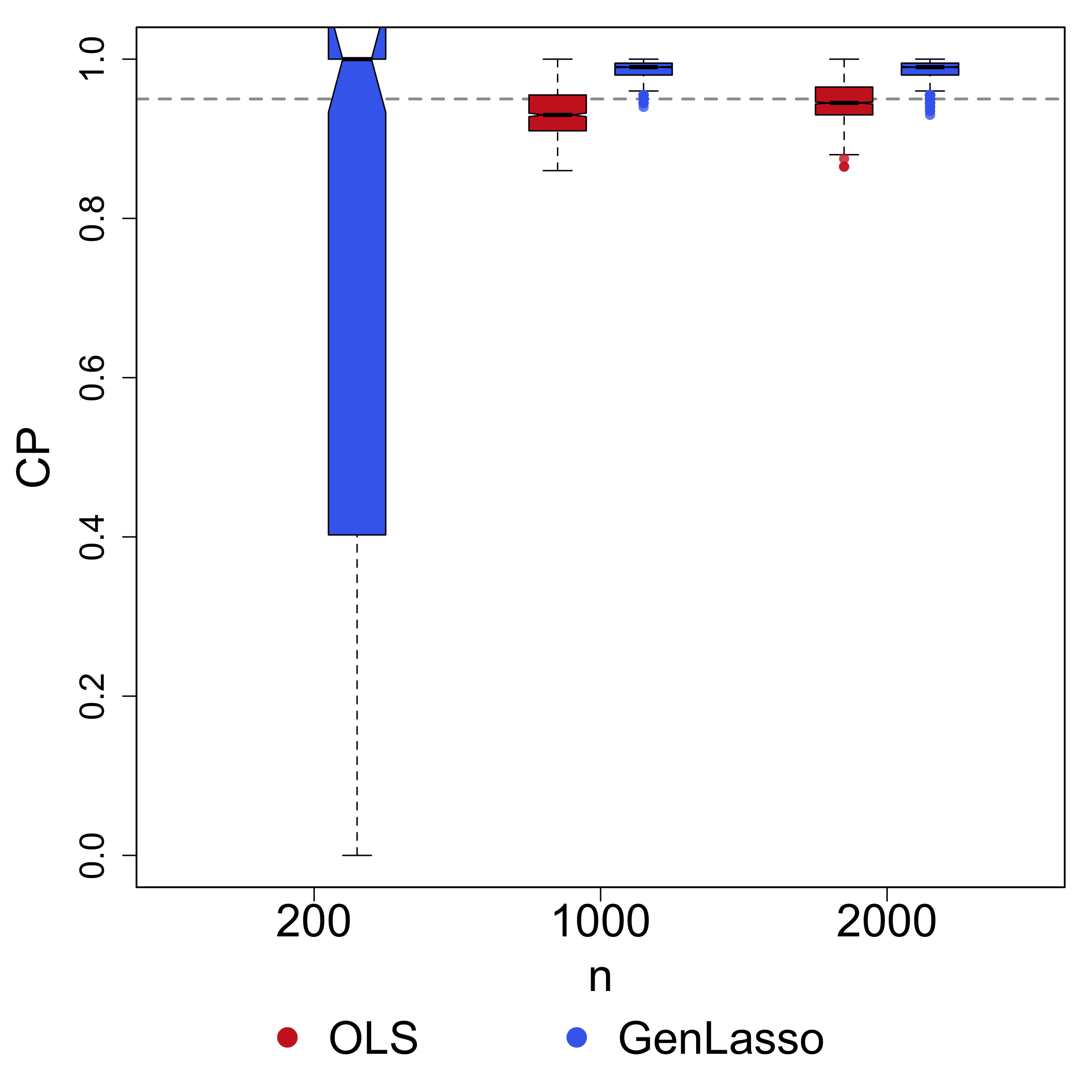}}
    \end{center}
    \caption{\label{fig:sim_inf} \small{Simulation studies: The scaled confidence interval (CI) size and coverage probability (CP) of cAIE and cADE from the OLS and GenLasso methods with sample sizes $n=200,1000,2000$ and $p=100$. Results are based on $200$ simulation replications.}}
\end{figure}

Figure~\ref{fig:sim_inf} shows expected behaviors for interval estimates: For OLS, Wald intervals achieve nominal coverage of $95\%$ when feasible, while for GenLasso, the multi-sample splitting procedure yields intervals with coverage approaching $95\%$ as the sample size increases.

Overall, the simulation study confirms that (1) the methods produce consistent estimates, (2) inference is reliable with sample splitting, and (3) the methods are practical for moderate-to-high dimensional mediation problems, under appropriate sample sizes.

\section{Re-analysis of the SPRINT trial data}
\label{sec:sprint}

We reanalyze the SPRINT data, which motivate the current study of heterogeneity in mediation effects. In this analysis, we use the proposed method to investigate whether early changes in UACR mediate the BP benefit induced by the SPRINT intervention in individual participants. Such mediation effects, if confirmed, will implicate albumin reduction as a general mechanism for BP lowering, regardless of drugs used. The initial evidence from classical CMA is conflicting, as shown in Table~\ref{table:sprint_cma}.

By expressing the mediation effects as functions of patient characteristics under LSEMs, cAIE and cADE are obtained as functions of patient characteristics. In lower-dimensional cases, it has been shown that the OLS estimates yield excellent performance. In high-dimensional situations, estimation can be achieved with the generalized lasso regularization. Together, the OLS and GenLasso methods represent an analytic strategy capable of identifying individuals for whom a given biological pathway is operative, and thus represent a step forward to more precise treatment tailoring.

Given the current sample size of SPRINT trial participants with renal impairment ($n=1,963$) and the number of pre-treatment covariates {($p=87$)}, we choose to use the OLS method with inference based on the asymptotic confidence intervals (Theorem~\ref{thm:OLS_asmp}).

\subsection{Analytical observations}

\begin{figure}
    \begin{center}
        {\includegraphics[width=0.80\textwidth]{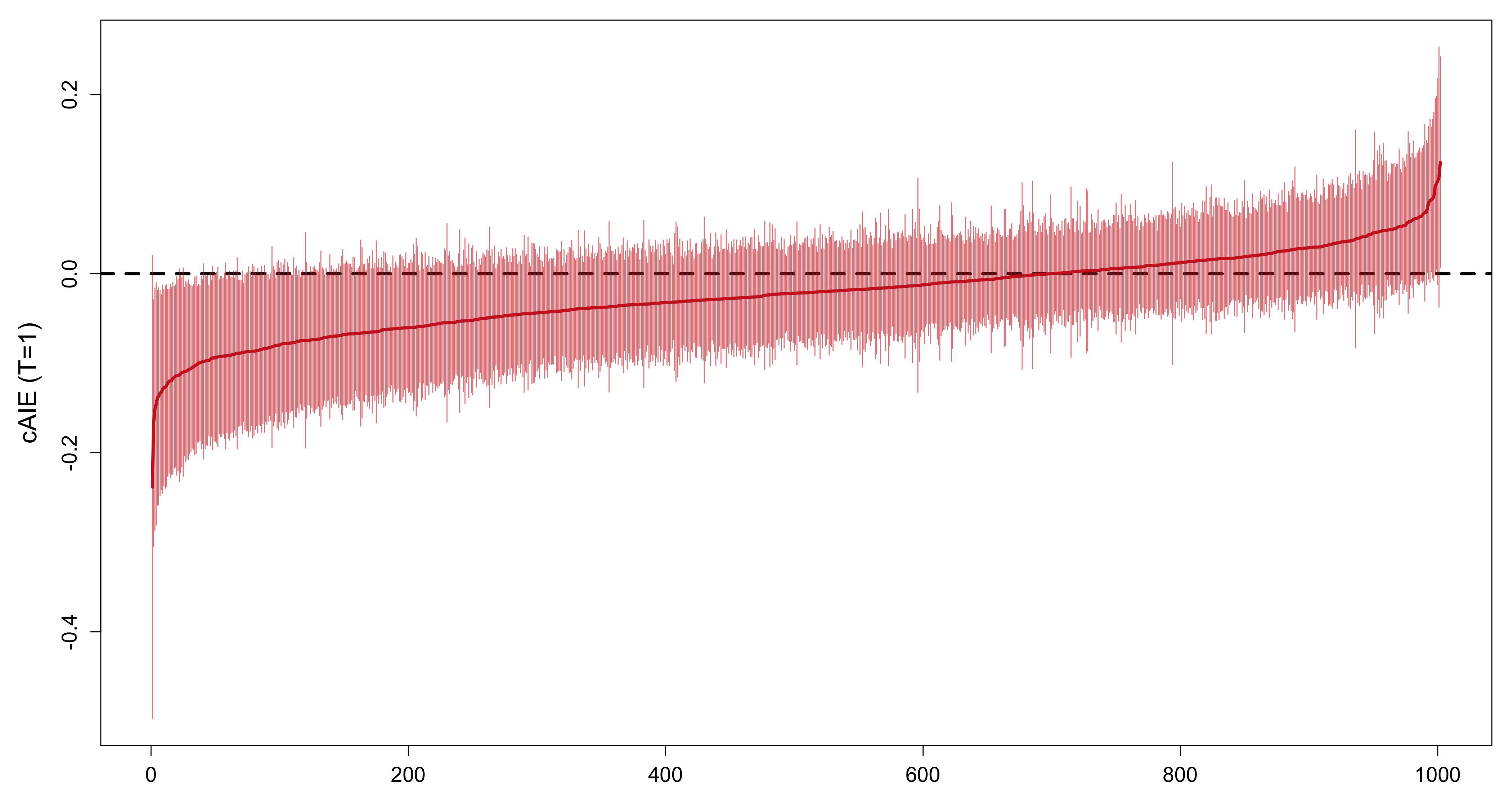}}
    \end{center}
    \caption{\label{fig:sprint}\small{SPRINT data analysis: Estimated conditional average indirect effect (cAIE) in patients who received the SPRINT intervention ($n=1002$). Each vertical line indicates the estimated cAIE and associated $95\%$ confidence interval (CI) of an individual patient. Estimates are calculated using the ordinary least squares (OLS) method, with CI calculated from the asymptotic distribution.}}
\end{figure}

We first estimate cAIEs associated with UACR reduction. The point and $95\%$ confidence interval estimates of cAIEs in the $1002$ participants with renal impairment who received the SPRINT intervention are presented in Figure~\ref{fig:sprint}. The analysis reveals substantial variability across individuals. For most SPRINT intervention recipients, the cAIE estimates are near zero, indicating that UACR changes do not contribute meaningfully to their BP reduction. Thus, as a whole, the data do not support the existence of a predominant UACR-related mechanism that has mediated the BP-lowering effect of the SPRINT intervention.

A distinct subgroup, however, shows a non-negligible mediation effect consistent with the UACR pathway. For this subgroup (shown at the left end of Figure~\ref{fig:sprint}), the UACR pathway appears to be active, with $95\%$ pointwise confidence intervals showing significant differences from zero. It is important to note that all of the significant cAIEs are on the negative side, meaning reduction of UACR at $6$ months mediates the treatment effects in lowering BP at $12$ months. When the estimated mediation effects are averaged across the broader group of participants with mild renal impairment, the accumulated signal produces a small but statistically significant negative population-average indirect effect; see Table~\ref{table:sprint_cma}. The new analysis demonstrates that this ``significant'' population-average mediation effect is almost certainly driven by a small subset of patients who mechanistically benefit through this specific pathway.

\begin{figure}
    \begin{center}
        \subfloat[\label{subfig:sprint_diff}]{\includegraphics[width=0.7\textwidth]{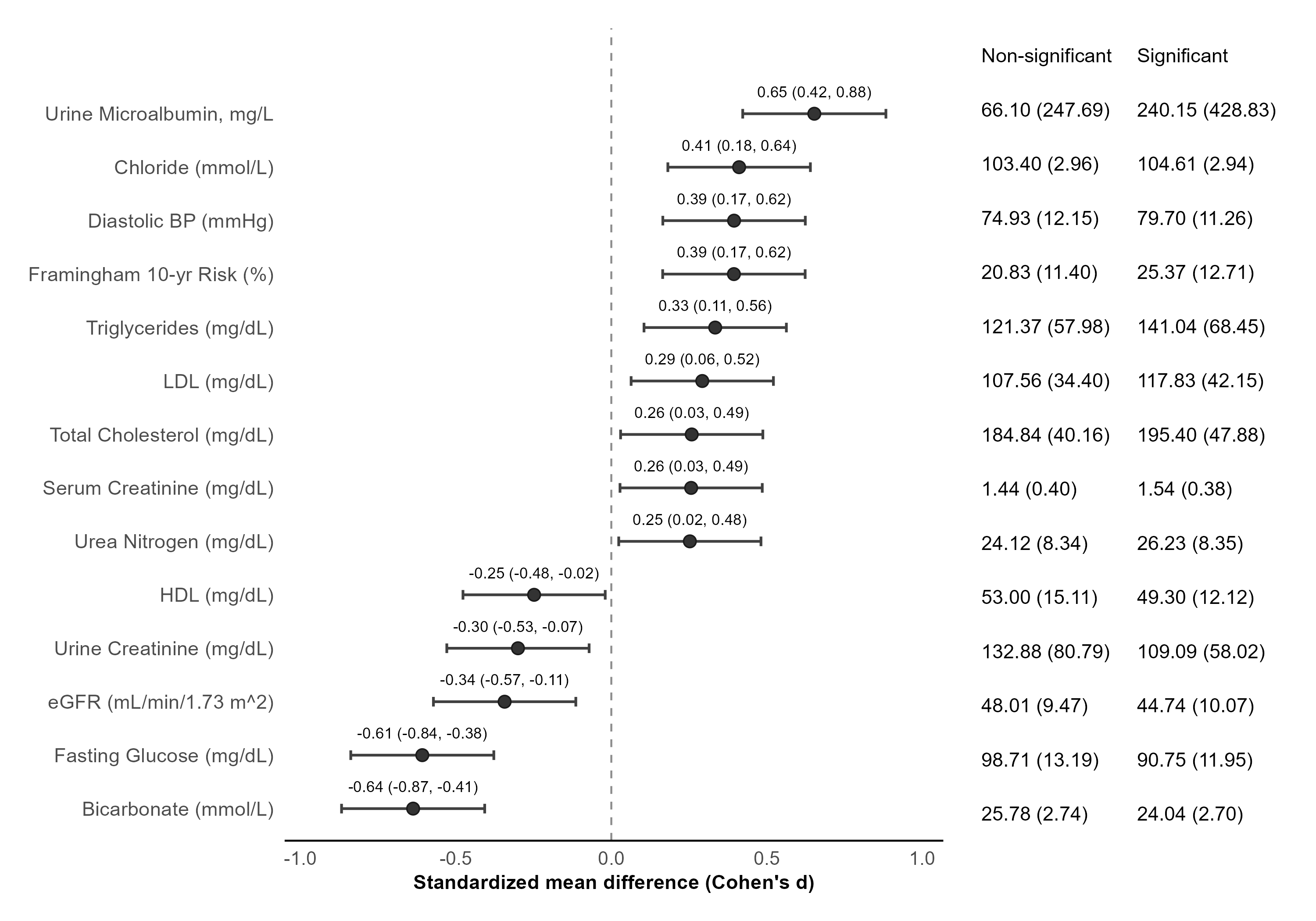}}
    
        \subfloat[\label{subfig:sprint_OR}]{\includegraphics[width=0.7\textwidth]{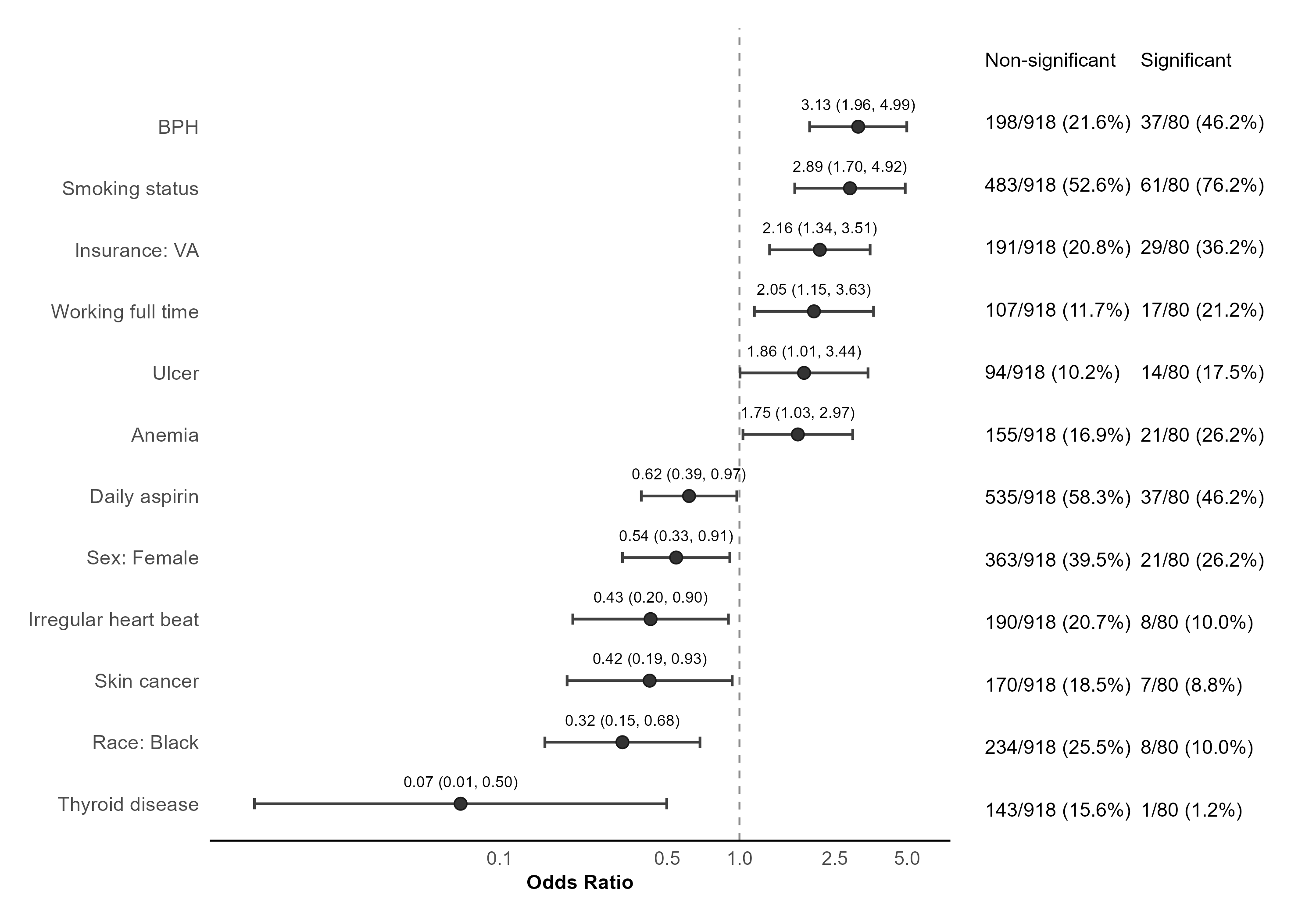}}
        
    \end{center}
    \caption{\label{fig:sprint-or} \small{SPRINT data analysis: Comparison of patient characteristics of the subgroups with a significant and non-significant conditional average indirect effect (cAIE). (a) Standardized mean difference (Cohen's $d$, and $95\%$ CI) between groups with non-significant cAIE versus significant cAIE for selected continuous covariates. (b) Odds ratio (OR, and $95\%$ CI) associated with having significant cAIE for selected categorical covariates. }}
\end{figure}

It is then logical to ask who the people are that benefit from UACR reduction. To answer this question, we further divide the SPRINT intervention recipients into two subgroups: Those who have a significant cAIE and those who do not. Subject characteristics are compared between these two newly defined groups. Variables that are statistically different are presented in Figure~\ref{fig:sprint-or}, where Panel~(a) contains the numerical variables and Panel~(b) contains the categorical variables.

\subsection{What do these findings mean clinically?}

Figure~\ref{subfig:sprint_diff} shows that patients with a significant cAIE through UACR reduction have a distinct clinical and biochemical profile, characterized by higher baseline albuminuria, worse kidney function (higher serum creatinine and urea nitrogen, lower eGFR and urine creatinine), mild metabolic acidosis (lower bicarbonate), and an adverse cardiovascular and metabolic risk profile (higher Framingham score, LDL cholesterol, triglycerides, and total cholesterol, with lower HDL). This constellation is consistent with a recognized ``high-risk vascular-renal phenotype'' in which albuminuria reflects active endothelial dysfunction and microvascular injury, not just a static biomarker.

These findings align closely with recent analysis of large-cohort and meta data showing that elevated albuminuria combined with moderate eGFR impairment identifies individuals at disproportionately high risk of cardiovascular events and CKD progression~\citep{heerspink2025proteinuria}. Poorly controlled BP in patients of this phenotype may simply reflect the worsening of their cardio-metabolic syndrome. Albuminuria is well established as a sensitive marker of systemic endothelial dysfunction and microvascular injury, particularly in patients with dyslipidemia and elevated atherosclerotic risk~\citep{claudel2025albuminuria}.

Interestingly, the categorical characteristics (Figure~\ref{subfig:sprint_OR}) show that the subgroup with significant cAIEs contains fewer Black and female participants; here, benign prostate hyperplasia (BPH) and health insurance covered by the Veteran's Affair Administration (VA) are likely markers for being male. Unlike the biochemical markers in Figure~\ref{subfig:sprint_diff}, these associations should be interpreted with caution -- as descriptive rather than mechanistic. Race and sex likely act as proxies for underlying biological or socio-environmental factors rather than being causal determinants themselves. Prior studies described the genetic basis for salt-sensitive HTN in  individuals of African ancestry~\citep{tu2013consideration}, but no studies to our knowledge have demonstrated UACR's mediation roles differed by sex or race. Daily aspirin use is less common among the benefiting subgroup. Similarly, while low-dose aspirin is known to be safe with no direct effects on albuminuria~\citep{taliercio2022aspirin}, daily aspirin use is likely a marker for certain health maintenance behavior rather than a mechanistic contributor.

In summary, the profile of patients who benefit from UACR reduction is clinically coherent: Individuals with greater renal impairment, higher albuminuria, and elevated vascular risk constitute this small subgroup in whom improvements in albuminuria likely represent real recovery of renal microvascular integrity. This explains why the UACR-mediated pathway appears strong in this subset but diminishes when averaged across the broader SPRINT population, where most participants do not fit this phenotype.

\subsection{Methodological takeaways}

This case study highlights a key point: Average mediation effects can be driven by a small set of individuals in whom the hypothesized biological mechanism is present and strong. This is the case in the SPRINT analysis, where the UACR mediation effect appears restricted to those with measurable glomerular injury, in whom intensive treatment yields BP improvement. From a physiological standpoint, this aligns well with the role of albuminuria as an early indicator of endothelial and podocyte dysfunction. Patients with higher baseline UACR with reversible glomerular injuries are more likely to respond to the intensive therapy; consequently, they are also more likely to experience reductions in albumin excretion that correlate with improved outcomes.

The heterogeneous estimates clarify why the population-average mediation analyses have produced inconsistent findings. Methods assuming homogeneity tend to dilute the strong renal-mediated signal present in a minority subgroup with the near-zero effect present in the majority. Depending on model specification, covariate adjustment, and sample composition, the population-average estimate may appear statistically significant, marginal, or null. Our method resolves the issue by identifying which patients actually exhibit a mediator-responsive pathway and quantifying its contribution on a personalized basis.

\section{Discussion}
\label{s:discuss}

This work introduces a framework for estimating heterogeneous mediation effects in settings where biological mechanisms vary with patient characteristics. Traditional CMA has largely focused on population-average mediation effects, implicitly assuming that the mediator is operational in all individuals. In many biomedical applications, this assumption is unlikely to hold. By expressing mediation effects as functions of covariates, the proposed method allows investigators to quantify mechanistic heterogeneity directly and identify the subset of individuals for whom the specific mediation pathway is active.

Methodologically, the framework extends classical CMA in two important ways. First, it accommodates interactions between covariates, treatment, and mediator, enabling the estimation of cADE and cAIE. Second, the modified covariate formulation relaxes the hierarchical constraints that complicate interaction modeling and better integrates with penalized regression. Coupled with the generalized lasso and a multiple sample-splitting inference procedure, the method is applicable in studies with many covariates while preserving model interpretability. These features make it suitable for modern clinical investigations where high-dimensional covariates have become the rule rather than the exception.

The LSEM-based method we propose is complementary to the recent development of nonparametric Bayesian approaches to heterogeneous mediation modeling~\citep{ting2025estimating}, offering an analytically transparent alternative that yields closed-form, covariate-dependent mediation contrasts. By relying on parametric structures, the method provides interpretable pathways and theoretically grounded inference even in the presence of high-dimensional covariates. These features make our method well suited for biomedical studies where mechanistic explanation and subgroup-specific effect decomposition are central.

The SPRINT data analysis shows how population-average CMA can mislead by masking important mechanistic variability. Although the average indirect effect through albuminuria reduction is small, the individualized estimates reveal a distinct subgroup, characterized by greater renal impairment, higher albuminuria, and more adverse cardiovascular profiles, in whom UACR reduction appears to mediate a meaningful portion of the BP benefit. Among the majority of participants, however, the estimated mediation effects are essentially zero, suggesting that alternative physiological pathways predominate. This pattern clarifies why standard analyses based on population averages produce conflicting findings: When mechanistic activity is confined to a minority, averaging distorts the true signal.

The classical formulation of NIE depends on cross-world independence assumptions that are inherently untestable~\citep{vanderweele2015explanation,andrews2021insights}. In the present work, identification is obtained under single-world sequential ignorability within a parametric linear structural equation framework, which avoids the need for cross-world assumptions but does not eliminate the possibility of unmeasured mediator–outcome confounding. The implications of such violations in heterogeneous settings merit further study, and future methodological work should incorporate sensitivity analyses or consider alternative estimands such as interventional indirect effects.


\section*{Data Availability}
{The SPRINT trial data are available with approval from the National Heart, Lung, and Blood Institute's (NHLBI) Biologic Specimen and Data Repository Information Coordinating Center, under signed NHLBI Research Materials Distribution Agreement (RMDA).}


\appendix
\counterwithin{figure}{section}
\counterwithin{table}{section}
\counterwithin{equation}{section}
\counterwithin{lemma}{section}
\counterwithin{theorem}{section}

This supplement has two sections: Section~\ref{appendix:sec:theory} contains proofs of the theoretical results. Section~\ref{appendix:sec:sprint} contains detailed information about participants in the SPRINT study, including a full list of characteristics of the SPRINT trial participants with renal impairment.


\section{Theory and proof}
\label{appendix:sec:theory}

\subsection{Proof of Theorem~\ref{thm:causal_def}}
\label{appendix:sub:proof_causal_def}

\begin{proof}
    To prove the identification of the causal effects, we have
    \begin{eqnarray*}
        \mathbb{E}(Y \mid T=t, M=m, \bZ=\bz) &=& \mathbb{E}(Y(t,m) \mid T=t, M=m, \bZ=\bz) \\
        &=& \mathbb{E}(Y(t,m) \mid T=t, M(t)=m, \bZ=\bz) \\
        &=& \mathbb{E}(Y(t,m) \mid T=t, \bZ=\bz) \\
        &=& \mathbb{E}(Y(t,m) \mid \bZ=\bz).
    \end{eqnarray*}
    From Line~4 to Line~3, the treatment ignorability assumption (A4) is applied, and from Line~3 to Line~2, the mediator ignorability assumption (A5) is utilized. From Line~2 to Line~1 and the equation in Line~1, the positivity (A2) and consistency (A3) assumptions are used. 
    Analogously, we have
    \[
        \mathbb{E}(M \mid T=t,\bZ=\bz) = \mathbb{E}(M(t) \mid T=t,\bZ=\bz) =\mathbb{E}(M(t) \mid \bZ=\bz),
    \]
    where the first equation uses the positivity and consistency assumption and the second equation uses the treatment ignorability assumption.
    Then, the causal effects can be identified from observed data.

    With the additional parametric model assumption (A6), one can show the parametric form of the causal effects as presented in Theorem~\ref{thm:causal_def}.
\end{proof}

\subsection{Proof of Theorem~\ref{thm:OLS_asmp}}
\label{appendix:sub:proof_OLS_asmp}

\begin{proof}
    Given the causal models~\eqref{eq:model_m} and~\eqref{eq:model_y}, for subject $i$ ($i=1,\dots,n$), models for the observed data are
    \begin{eqnarray*}
      M_{i} &=& \balpha_{0}^\top\bZ_{i}+\balpha_{1}^\top\bZ_{i} T_{i}+\varepsilon_{i}, \\
      Y_{i} &=& \bgamma_{0}^\top\bZ_{i}+\bgamma_{1}^\top\bZ_{i} T_{i}+\beta_{0}M_{i}+\beta_{1}M_{i}T_{i}+\eta_{i}.
    \end{eqnarray*}
    Let
     \[
      \bZ=\begin{pmatrix}
       \bZ_{1}^\top \\
       \vdots \\
       \bZ_{n}^\top
      \end{pmatrix}_{n\times p}, ~\bM=\begin{pmatrix}
       M_{1} \\
       \vdots \\
       M_{n}
      \end{pmatrix}_{n\times 1}, ~\bY=\begin{pmatrix}
       Y_{1} \\
       \vdots \\
       Y_{n}
      \end{pmatrix}_{n\times 1}, ~\bT=\diag\{T_{1},\dots,T_{n}\}\in\mathbb{R}^{n\times n},
     \]
    and define
    \[
      \bU=\bT\bZ=\begin{pmatrix}
       T_{1}\bZ_{1}^\top \\
       \vdots \\
       T_{n}\bZ_{n}^\top
      \end{pmatrix}_{n\times p}, ~\bV=\bT\bM=\begin{pmatrix}
       T_{1}M_{1} \\
       \vdots \\
       T_{n}M_{n}
      \end{pmatrix}_{n\times 1}, ~\bvarepsilon=\begin{pmatrix}
       \varepsilon_{1} \\
       \vdots \\
       \varepsilon_{n}
      \end{pmatrix}_{n\times 1}, ~\bdeta=\begin{pmatrix}
       \eta_{1} \\
       \vdots \\
       \eta_{n}
      \end{pmatrix}_{n\times 1}.
     \]
     For $n$ subjects, the models can be written in the following form,
     \begin{eqnarray*}
      \bM &=& \bZ\balpha_{0}+\bU\balpha_{1}+\bvarepsilon, \\
      \bY &=& \bZ\bgamma_{0}+\bU\bgamma_{1}+\bM\beta_{0}+\bV\beta_{1}+\bdeta,
     \end{eqnarray*}
     where
     \[
      \bvarepsilon\sim F(\matzero,\sigma_{m}^{2}\matI_{n}), \quad \bdeta\sim F(\matzero,\sigma_{y}^{2}\matI_{n}), \quad \text{and }~ \bvarepsilon~\indep~\bdeta.
     \]
     Combine the two models,
     \[
      \begin{pmatrix}
       \bM & \bY
      \end{pmatrix}_{n\times 2}=\begin{pmatrix}
       \bZ & \bU & \bM & \bV
      \end{pmatrix}_{n\times(2p+2)}\begin{pmatrix}
       \balpha_{0} & \bgamma_{0} \\
       \balpha_{1} & \bgamma_{1} \\
       0 & \beta_{0} \\
       0 & \beta_{1}
      \end{pmatrix}_{(2p+2)\times 2}+\begin{pmatrix}
       \bvarepsilon & \bdeta
      \end{pmatrix}_{n\times 2},
     \]
     and rewrite it as
     \[
      \bO=\bX\bTheta+\bzeta,
     \]
     where
     \[
        \bO=\begin{pmatrix}
       \bM & \bY
      \end{pmatrix}_{n\times 2}, ~ \bX=\begin{pmatrix}
       \bZ & \bU & \bM & \bV
      \end{pmatrix}_{n\times(2p+2)}, ~ \bTheta=\begin{pmatrix}
       \balpha_{0} & \bgamma_{0} \\
       \balpha_{1} & \bgamma_{1} \\
       0 & \beta_{0} \\
       0 & \beta_{1}
      \end{pmatrix}_{(2p+2)\times 2}, ~ \bzeta=\begin{pmatrix}
       \bvarepsilon & \bdeta
      \end{pmatrix}_{n\times 2},
     \]
     \[
      \bzeta_{i}\sim F(\matzero,\bSigma), \quad \bSigma=\begin{pmatrix}
       \sigma_{m}^{2} & 0 \\
       0 & \sigma_{y}^{2}
      \end{pmatrix}.
     \]
     The ordinary least squares (OLS) estimator of $\bTheta$ is
     \[
      \hat{\bTheta}=(\bX^\top\bX)^{-1}\bX^\top\bO=\bTheta+(\bX^\top\bX)^{-1}\bX^\top\bzeta.
     \]
     \begin{eqnarray*}
      \sqrt{n}\mathrm{vec}(\hat{\bTheta}-\bTheta) &=& \mathrm{vec}\left\{\left(\frac{1}{n}\bX^\top\bX \right)^{-1}\left(\frac{1}{\sqrt{n}}\bX^\top\bzeta \right)\right\} \\
      &=& \matI_{2}\otimes \left(\frac{1}{n}\bX^\top\bX \right)^{-1} \mathrm{vec}\left(\frac{1}{\sqrt{n}}\bX^\top\bzeta \right),
     \end{eqnarray*}
     using the fact that $\mathrm{vec}(\bA\bB\bC)=(\bC^\top\otimes\bA)\mathrm{vec}(\bB)$ for matrices $\bA$, $\bB$, and $\bC$.
     To derive the asymptotic distribution of $\hat{\bTheta}$, we first study the asymptotics of $(\bX^\top\bX/n)$.
     \[
      \bX^\top\bX = \begin{pmatrix}
       \bZ^\top \\
       \bU^\top \\
       \bM^\top \\
       \bV^\top
      \end{pmatrix}\begin{pmatrix}
       \bZ & \bU & \bM & \bV
      \end{pmatrix} =\begin{pmatrix}
       \bZ^\top\bZ & \bZ^\top\bU & \bZ^\top\bM & \bZ^\top\bV \\
       \bU^\top\bZ & \bU^\top\bU & \bU^\top\bM & \bU^\top\bV \\
       \bM^\top\bZ & \bM^\top\bU & \bM^\top\bM & \bM^\top\bV \\
       \bV^\top\bZ & \bV^\top\bU & \bV^\top\bM & \bV^\top\bV
      \end{pmatrix}.
     \]
     Assume $\bZ^\top\bZ/n\rightarrow \bQ\in\mathbb{R}^{p\times p}$, $\mathbb{E}(T_{i})=\pi$, $\mathbb{E}(T_{i}^{2})=\pi_{2}$, $\mathbb{E}(T_{i}^{3})=\pi_{3}$, $\mathbb{E}(T_{i}^{4})=\pi_{4}$. Then,
     \[
      \frac{1}{n}\bZ^\top\bU=\frac{1}{n}\bZ^\top\bT\bZ\rightarrow \pi\bQ, \quad \frac{1}{n}\bU^\top\bU=\frac{1}{n}\bZ^\top\bT^{2}\bZ\rightarrow\pi_{2}\bQ,
     \]
     \[
      \frac{1}{n}\bZ^\top\bM=\frac{1}{n}\left(\bZ^\top\bZ\balpha_{0}+\bZ^\top\bU\balpha_{1}+\bZ^\top\bvarepsilon \right)\rightarrow \bQ\balpha_{0}+\pi\bQ\balpha_{1}\coloneqq \bQ_{ZM},
     \]
     \[
      \frac{1}{n}\bZ^\top\bV=\frac{1}{n}\left(\bZ^\top\bT\bZ\balpha_{0}+\bZ^\top\bT^{2}\bZ\balpha_{1}+\bZ^\top\bT\bvarepsilon \right)\rightarrow \pi\bQ\balpha_{0}+\pi_{2}\bQ\balpha_{1}\coloneqq \bQ_{ZV}.
     \]
     Here $\bZ^\top\bvarepsilon/n\rightarrow \matzero$ and $\bZ^\top\bT\bvarepsilon/n\rightarrow\matzero$ due to independence.
     \[
      \frac{1}{n}\bU^\top\bM=\frac{1}{n}\left(\bZ^\top\bT\bZ\balpha_{0}+\bZ^\top\bT^{2}\bZ\balpha_{1}+\bZ^\top\bT\bvarepsilon \right)\rightarrow \pi\bQ\balpha_{0}+\pi_{2}\bQ\balpha_{1}\coloneqq \bQ_{UM},
     \]
     \[
      \frac{1}{n}\bU^\top\bV=\frac{1}{n}\left(\bZ^\top\bT^{2}\bZ\balpha_{0}+\bZ^\top\bT^{3}\bZ\balpha_{1}+\bZ^\top\bT^{2}\bvarepsilon \right)\rightarrow \pi_{2}\bQ\balpha_{0}+\pi_{3}\bQ\balpha_{1}\coloneqq \bQ_{UV},
     \]
     \begin{eqnarray*}
      \frac{1}{n}\bM^\top\bM &=& \frac{1}{n}\left(\balpha_{0}^\top\bZ^\top\bZ\balpha_{0}+\balpha_{0}^\top\bZ^\top\bU\balpha_{1}+\balpha_{0}^\top\bZ^\top\bvarepsilon+\balpha_{1}^\top\bU^\top\bZ\balpha_{0} \right. \\
      && \quad \quad \left. +\balpha_{1}^\top\bU^\top\bU\balpha_{1}+\balpha_{1}^\top\bU^\top\bvarepsilon+\bvarepsilon^\top\bZ\balpha_{0}+\bvarepsilon^\top\bU\balpha_{1}+\bvarepsilon^\top\bvarepsilon \right) \\
      &\rightarrow& \balpha_{0}^\top\bQ\balpha_{0}+2\pi\balpha_{0}^\top\bQ\balpha_{1}+\pi_{2}\balpha_{1}^\top\bQ\balpha_{1}+\sigma_{m}^{2} \\
      &\coloneqq& \bQ_{M},
     \end{eqnarray*} 
     \begin{eqnarray*}
      \frac{1}{n}\bM^\top\bV &=& \frac{1}{n}\left(\balpha_{0}^\top\bZ^\top\bT\bZ\balpha_{0}+\balpha_{0}^\top\bZ^\top\bT\bU\balpha_{1}+\balpha_{0}^\top\bZ^\top\bT\bvarepsilon+\balpha_{1}^\top\bU^\top\bT\bZ\balpha_{0} \right. \\
      && \quad \quad \left. +\balpha_{1}^\top\bU^\top\bT\bU\balpha_{1}+\balpha_{1}^\top\bU^\top\bT\bvarepsilon+\bvarepsilon^\top\bT\bZ\balpha_{0}+\bvarepsilon^\top\bT\bU\balpha_{1}+\bvarepsilon^\top\bT\bvarepsilon \right) \\
      &\rightarrow& \pi\balpha_{0}^\top\bQ\balpha_{0}+2\pi_{2}\balpha_{0}^\top\bQ\balpha_{1}+\pi_{3}\balpha_{1}^\top\bQ\balpha_{1}+\pi\sigma_{m}^{2} \\
      &\coloneqq& \bQ_{MV},
     \end{eqnarray*} 
     \begin{eqnarray*}
      \frac{1}{n}\bV^\top\bV &=& \frac{1}{n}\left(\balpha_{0}^\top\bZ^\top\bT^{2}\bZ\balpha_{0}+\balpha_{0}^\top\bZ^\top\bT^{2}\bU\balpha_{1}+\balpha_{0}^\top\bZ^\top\bT^{2}\bvarepsilon+\balpha_{1}^\top\bU^\top\bT^{2}\bZ\balpha_{0} \right. \\
      && \quad \quad \left. +\balpha_{1}^\top\bU^\top\bT^{2}\bU\balpha_{1}+\balpha_{1}^\top\bU^\top\bT^{2}\bvarepsilon+\bvarepsilon^\top\bT^{2}\bZ\balpha_{0}+\bvarepsilon^\top\bT^{2}\bU\balpha_{1}+\bvarepsilon^\top\bT^{2}\bvarepsilon \right) \\
      &\rightarrow& \pi_{2}\balpha_{0}^\top\bQ\balpha_{0}+2\pi_{3}\balpha_{0}^\top\bQ\balpha_{1}+\pi_{4}\balpha_{1}^\top\bQ\balpha_{1}+\pi_{2}\sigma_{m}^{2} \\
      &\coloneqq& \bQ_{V}.
     \end{eqnarray*}
     Thus,
     \[
      \frac{1}{n}\bX^\top\bX \rightarrow \begin{pmatrix}
       \bQ & \pi\bQ & \bQ_{ZM} & \bQ_{ZV} \\
       \pi\bQ & \pi_{2}\bQ & \bQ_{UM} & \bQ_{UV} \\
       \bQ_{ZM} & \bQ_{UM} & \bQ_{M} & \bQ_{MV} \\
       \bQ_{ZV} & \bQ_{UV} & \bQ_{MV} & \bQ_{V}
      \end{pmatrix}\coloneqq ~\bQ_{X}
     \]
     Using the relation $\mathrm{vec}(\bA\bB)=(\bB^\top\otimes\bA)\mathrm{vec}(\matI)$,
     \[
      \frac{1}{\sqrt{n}}\mathrm{vec}(\bX^\top\bzeta)=\frac{1}{\sqrt{n}}\sum_{i=1}^{n}\mathrm{vec}(\bX_{i}\bzeta_{i}^\top)=\frac{1}{\sqrt{n}}\sum_{i=1}^{n}\bzeta_{i}\otimes \bX_{i}.
     \]
     \[
      \Var(\bzeta_{i}\otimes\bX_{i})=\mathbb{E}\left\{(\bzeta_{i}\bzeta_{i}^\top)\otimes (\bX_{i}\bX_{i}^\top) \right\} \quad \Rightarrow \quad \frac{1}{\sqrt{n}}\sum_{i=1}^{n}\bzeta_{i}\otimes\bX_{i} \overset{\mathcal{D}}{\longrightarrow} \mathcal{N}(\matzero,\bXi), \quad \bXi=\bSigma\otimes \bQ_{X}.
     \]

     Now consider the asymptotic distribution of cAIE and cADE. For given $t$ and $\bz$,
 \begin{eqnarray*}
  \delta(t\mid \bz) &=& 2(\beta_{0}+\beta_{1}t)\balpha_{1}^\top\bz \coloneqq f_{1}(\bTheta), \\
  \xi(t\mid \bz) &=& 2\bgamma_{1}^\top\bz+2\beta_{1}(\balpha_{0}^\top\bz+\balpha_{1}^\top\bz t) \coloneqq f_{2}(\bTheta).
 \end{eqnarray*}
 \[
  \frac{\partial f_{1}}{\partial\mathrm{vec}(\bTheta)}=\begin{pmatrix}
   \matzero \\
   2(\beta_{0}+\beta_{1}t)\bz \\
   0 \\
   0 \\
   \matzero \\
   \matzero \\
   2\balpha_{1}^\top\bz \\
   2t\balpha_{1}^\top\bz
  \end{pmatrix}\coloneqq \bH_{1}, \quad \frac{\partial f_{2}}{\partial\mathrm{vec}(\bTheta)}=\begin{pmatrix}
   2\beta_{1}\bz \\
   2\beta_{1}t\bz \\
   0 \\
   0 \\
   \matzero \\
   2\bz \\
   0 \\
   2(\balpha_{0}^\top\bz+\balpha_{1}^\top\bz t) 
  \end{pmatrix}\coloneqq \bH_{2}.
 \]
 Using the multivariate Delta method,
 \[
  \sqrt{n}(\hat{\delta}-\delta)\mid t,\bz \overset{\mathcal{D}}{\longrightarrow}\mathcal{N}\left(0, \bH_{1}^\top (\bSigma\otimes\bQ_{X}^{-1})\bH_{1} \right)
 \]
 \[
  \sqrt{n}(\hat{\xi}-\xi)\mid t,\bz \overset{\mathcal{D}}{\longrightarrow}\mathcal{N}\left(0, \bH_{2}^\top (\bSigma\otimes\bQ_{X}^{-1})\bH_{2} \right)
 \]
\end{proof}

\subsection{Consistency of generalized lasso}
\label{appendix:sub:genlasso_asmp}

This section discusses the consistency of the generalized lasso estimator. Following the notation in Section~\ref{sub:method}, where a generic form of the models is rewritten as
\begin{equation*}
    \tilde{\bR}=\tilde{\bS}\bphi+\tilde{\bvarepsilon}.
\end{equation*}
And consider the following generalized lasso estimator with regularization matrix $\bD$,
\begin{equation*}
    \underset{\bphi}{\text{minimize}} ~\frac{1}{2}\|\tilde{\bR}-\tilde{\bS}\bphi\|_{2}^{2}+\lambda\|\bD\bphi\|_{1}.
\end{equation*}
Let $\bphi^{*}$ denote the true model parameter and $\mathscr{M}=\{\bphi\in\mathbb{R}^{2q}\mid (\bD\bphi)_{\mathcal{S}^{c}}=\matzero \}$ denote the model space, where $\mathcal{S}$ is the support of $\bD\bphi$ and $\mathcal{S}^{c}$ is the complement.
Following \citet{lee2015model}, two regularity conditions are imposed.
\begin{description}
    \item[Condition 1] (Restricted strong convexity, RSC) Let $\mathcal{C}\subset\mathbb{R}^{2q}$ be a known convex set containing $\bphi^{*}$. The loss function $\ell=\|\tilde{\bR}-\tilde{\bS}\bphi\|_{2}^{2}/2$ is RSC on $\mathcal{C}\cap\mathscr{M}$ when
    \[
        \btheta^\top\nabla^{2}\ell(\bphi)\btheta\geq m\|\btheta\|_{2}^{2}, \quad \bphi\in\mathcal{C}\cap\mathscr{M}, \quad \btheta\in(\mathcal{C}\cap\mathscr{M})-(\mathcal{C}\cap\mathscr{M}),
    \]
    \[
        \|\nabla^{2}\ell(\bphi)-Q\|_{2}\leq L\|\bphi-\bphi^{*}\|_{2}, \quad \bphi\in\mathcal{C},
    \]
    for some $m>0$ and $L<\infty$, where $Q=\nabla^{2}\ell(\bphi^{*})$ is the sample Fisher information matrix and $\nabla$ is the differential operator.

    \item[Condition 2] For $\iota\in(0,1)$,
    \[
        \|\bD_{\mathcal{S}^{c}}\tilde{\bS}^\top(\bD_{\mathcal{S}^{c}}\tilde{\bS}^\top)^{-}\sign\{(\bD\bphi^{*})_{\mathcal{S}}\}\|_{\infty}\leq 1-\iota,
    \]
    where $\bD_{\mathcal{S}}\in\mathbb{R}^{|\mathcal{S}|\times 2q}$ takes the rows of $\bD$ in $\mathcal{S}$, $(\bD\bphi^{*})\in\mathbb{R}^{|\mathcal{S}|}$, and $\bA^{-}$ is the Moore-Penrose pseudoinverse of a matrix $\bA\in\mathbb{R}^{2q\times 2q}$ and $\sign(\cdot)$ is the sign function.
\end{description}
Both conditions are regularity conditions on the Fisher information matrix $Q$, where the second is a condition with respect to the regularization matrix $\bD$. 
With independent samples, Theorem~1 of \citet{raskutti2010restricted} implies that, with probability $1-c_{1}\exp\{-c_{2}n\}$, $\btheta^\top\nabla^{2}\ell(\bphi)\btheta\geq \left(\|\bSigma^{1/2}\btheta\|_{2}/4-9\rho(\bSigma)\sqrt{\log(2q)/n}\|\btheta\|_{1}\right)^{2}$, where $\bSigma$ is the covariance matrix of $\tilde{\bS}$, $\rho^{2}(\bSigma)=\max_{j}\Sigma_{jj}$ is the largest variance, and $c_{1},c_{2}$ are constants. It can be shown that with $m=\pi_{\min}/64$, this condition is satisfied when $n>4\cdot(36^{2}/\pi_{\min})\rho^{2}(\bSigma)|\mathcal{S}|\log(2q)$, where $\pi_{\min}$ is the smallest eigenvalue of $\bSigma$.
Condition~2 is a relaxation of the orthogonal design condition to near orthogonality.

In Theorem~\ref{thm:genlasso_asmp}, three compatibility constants, $\kappa_{1},\kappa_{2},\kappa_{3}$, are introduced. Let $\kappa_{\mathcal{R}}$ denote the compatibility constant between the regularization $\mathcal{R}(\bphi)=\|\bD\bphi\|_{1}$ and the $\ell_{2}$-norm on $\mathscr{M}$,
\[
    \kappa_{\mathcal{R}}=\sup_{\bphi}\left\{\mathcal{R}(\bphi):\bphi\in\mathscr{B}_{2}\cap\mathscr{M} \right\},
\]
where $\mathscr{B}_{2}=\{\bx\in\mathbb{R}^{2q}:\|\bx\|_{2}\leq 1\}$ is the $\ell_{2}$-norm ball. Let $\nu$ be some norm on $\mathbf{R}^{2q}$ such that $\mathcal{R}(\bphi)\leq \nu(\bphi)$ for any $\bphi\in\mathbb{R}^{2q}$. Use $\kappa_{\text{IC}}$ to denote the compatibility constant between the irrepresentable term and $\nu$,
\[
    \kappa_{\text{IC}}=\sup_{\nu(\bx)\leq 1} V\left[P_{\mathscr{M}^{\perp}}\left\{QP_{\mathscr{M}}(P_{\mathscr{M}}QP_{\mathscr{M}})^{-}P_{\mathscr{M}}\bx-\bx \right\} \right],
\]
where $P_{\mathscr{M}}\bx$ denotes the projection of $\bx$ on $\mathrm{span}(\mathscr{M})$. Below proves Theorem~\ref{thm:genlasso_asmp} by first providing the value of the compatibility constants $\kappa_{1},\kappa_{2},\kappa_{3}$.

\begin{proof}
    It can be shown that $\bD$ is invertible and has a nontrivial null space. Let $\nu=\|\cdot\|_{1}$, the compatibility constants are computed as the following:
    \[
        \kappa_{1}=\kappa_{\text{IC}}=\|\bD_{\mathcal{S}^{c}}\tilde{\bS}^\top(\bD_{\mathcal{S}^{c}}\tilde{\bS}^\top)^{-}\sign(\bphi_{\mathcal{S}}^{*})\|_{\infty},
    \]
    \[
        \kappa_{2}=\kappa_{\mathcal{R}}=\sup_{\bphi}\left\{\|\bD\bphi\|_{1}:\bphi\in\mathscr{B}_{2}\cap\mathrm{span}(\bD^\top\mathscr{B}_{\infty,\mathcal{S}^{c}})^{\perp} \right\},
    \]
    \[
        \kappa_{3}=\kappa_{\nu}=\sup_{\bphi}\left\{\|\bphi\|_{1}:\bphi\in\mathscr{B}_{2}\cap\mathrm{span}(\bD^\top\mathscr{B}_{\infty,\mathcal{S}^{c}})^{\perp} \right\}.
    \]
    $\mathcal{R}$ and $\nu$ are finite, $\kappa_{1},\kappa_{2},\kappa_{3}<\infty$. The rest of the proof can be found in \citet{lee2015model}.

    With the consistency of the estimate of model coefficients, the consistency of estimating cAIE and cADE with a given treatment assignment $t$ and a vector of covariates $\bz$ follows.
\end{proof}


\section{Additional data of the SPRINT study}
\label{appendix:sec:sprint}

\subsection{Characteristics of SPRINT trial participants}
\label{appendix:sub:sprint_charact}

In this section, we present the baseline characteristics of the SPRINT trial participants with renal impairment at the study entry, defined as baseline eGFR $\le 60$ mL/min/$1.73$ m$^2$. This results in a total of $n=1,963$ participants, $1,002$ of whom received the SPRINT intervention and $961$ received control treatment.

Considering the large number of patient characteristics in the current analysis, we group these characteristics into four major groups: Demographic and lifestyle factors, clinical factors, major comorbidities, and laboratory measures. 

\begin{table}[htbp]
    \begin{center}
        \caption{\label{appendix:table:sprint_demo}Demographic and lifestyle characteristics} 
        \resizebox{\textwidth}{!}{
        \begin{tabular}{p{8cm} >{\raggedleft\arraybackslash}p{3cm} >{\raggedleft\arraybackslash}p{3cm} >{\raggedleft\arraybackslash}p{2cm}}
        \toprule
        & \multicolumn{1}{c}{\textbf{Control}} & \multicolumn{1}{c}{\textbf{SPRINT}} \\
        \multicolumn{1}{c}{\multirow{-2}{*}{\textbf{Characteristic}}} & \multicolumn{1}{c}{($n=961$)} & \multicolumn{1}{c}{($n=1002$)} & \multicolumn{1}{c}{\multirow{-2}{*}{\textbf{$p$-value}}} \\
        \midrule
        
        Male sex& 597 (62.1\%) & 616 (61.5\%) & 0.768 \\
        
        Black race & 224 (23.3\%) & 245 (24.5\%) & 0.553\\
        
        Age $\ge 75$ years & 442 (43.9\%) & 442 (44.1\%) & 0.929 \\
        
        Live with other adults & 694 (72.2\%) & 714 (71.3\%) & 0.637\\
        
        Highest level of education completed? & & & 0.883 \\
        \quad College Graduate & 145 (15.1\%) & 153 (15.3\%) & \\
        \quad Grade School (5--8 years) & 24 (2.5\%) & 23 (2.3\%) & \\
        \quad Some High School (9--11 years) & 65 (6.8\%) & 67 (6.7\%) & \\
        \quad High School Diploma or G.E.D. & 172 (17.9\%) & 170 (17.0\%) & \\
        \quad Business/Vocational Training & 67 (7.0\%) & 76 (7.6\%) & \\
        \quad Some College (No degree) & 191 (19.9\%) & 205 (20.5\%) & \\
        \quad Associate Degree (A.D. or A.A.) & 47 (4.9\%) & 61 (6.1\%) & \\
        \quad Some Grad/Professional School & 78 (8.1\%) & 78 (7.8\%) & \\
        \quad Master’s Degree & 102 (10.6\%) & 101 (10.1\%) & \\
        \quad Doctoral Degree (Ph.D., M.D., J.D., etc.) & 56 (5.8\%) & 55 (5.5\%) & \\
        \quad Missing & 14 (1.5\%) & 13 (1.3\%) & \\
        
        Are you working full time for pay & 144 (15.0\%) & 125 (12.5\%) &0.106 \\
        
        \quad If No, are you retired & 669 (69.6\%) & 738 (73.7\%) & 0.047\\
        
        \quad If No, are you working part-time? & 134 (13.9\%) & 120 (12.0\%) & 0.194\\
        
        \quad If No, are you keeping house & 39 (4.1\%) & 61 (6.1\%) &0.041 \\
        
        \quad If No, are you unemployed or laid off & 52 (5.4\%) & 49 (4.9\%) & 0.602\\
        
        \quad If No, are you looking for work & 30 (3.1\%) & 25 (2.5\%) & 0.400 \\
        
        Insurance: Medicare & 663 (69.0\%) & 681 (68.0\%) & 0.625 \\
        
        Insurance: Medicaid & 81 (8.4\%) & 77 (7.7\%) & 0.545 \\
        
        Insurance: VA & 206 (21.4\%) & 220 (22.0\%) &  0.780\\
        
        Insurance: Private/Other & 393 (40.9\%) & 429 (42.8\%) & 0.389\\
        
        Insurance: Uninsured & 51 (5.3\%) & 57 (5.7\%) & 0.711\\
        
        Have co-pay for doctor’s or ER visits & 498 (51.8\%) & 531 (53.0\%) &0.603 \\
        
        Have to get a referral to see a specialist & 210 (21.9\%) & 204 (20.4\%) &0.418 \\
        
        Neither co-pay nor referrals required & 278 (28.9\%) & 290 (28.9\%) & 0.995 \\
        
        Insurance: Don’t know & 65 (6.8\%) & 59 (5.9\%) & 0.425\\
        
        \bottomrule
        \end{tabular}
        }
    \end{center}
\end{table}

\newpage

\addtocounter{table}{-1}
\begin{table}[t]
    \begin{center}
        \caption{Demographic and lifestyle characteristics (continued)}
        \resizebox{\textwidth}{!}{
        \begin{tabular}{p{8cm} >{\raggedleft\arraybackslash}p{3cm} >{\raggedleft\arraybackslash}p{3cm} >{\raggedleft\arraybackslash}p{2cm}}
        \toprule
        & \multicolumn{1}{c}{\textbf{Control}} & \multicolumn{1}{c}{\textbf{SPRINT}} \\
        \multicolumn{1}{c}{\multirow{-2}{*}{\textbf{Characteristic}}} & \multicolumn{1}{c}{($n=961$)} & \multicolumn{1}{c}{($n=1002$)} & \multicolumn{1}{c}{\multirow{-2}{*}{\textbf{$p$-value}}} \\
        \midrule
        
        Family history of heart disease, MI, stroke & 592 (61.6\%) & 647 (64.6\%) & 0.301\\
        
        Drank alcohol during the last 12 months & 576 (59.9\%) & 584 (58.3\%) & 0.456 \\
        
        Smoked at least 100 cigarettes in your life & 518 (53.9\%) & 532 (53.1\%) & 0.720\\
        
        Smoking status & 522 (54.3\%) & 546 (54.5\%) & 0.939  \\
        
        Frequency of vigorous activities (Mean (SD)) & 2.80 (1.48) & 2.71 (1.46) & 0.154 \\
        
        Time spent in less vigorous activities (Mean (SD)) & 2.57 (1.19) & 2.58 (1.19) & 0.889 \\
        
        Usual source of care & & & 0.213 \\
        \quad Private doctor’s office & 511 (53.2\%) & 516 (51.5\%) & \\
        \quad Hospital clinic or outpatient department & 343 (35.7\%) & 359 (35.8\%) & \\
        \quad Community health center & 45 (4.7\%) & 47 (4.7\%) & \\
        \quad Other kind of health care facility & 47 (4.9\%) & 60 (6.0\%) & \\
        \quad No usual source of care & 15 (1.6\%) & 20 (2.0\%) & \\
        
        \bottomrule
        \end{tabular}
        }
    \end{center}
\end{table}

\newpage

\begin{table}[htbp]
    \begin{center}
        \caption{\label{appendix:table:sprint_clinical}Clinical characteristics}
        \resizebox{\textwidth}{!}{
        \begin{tabular}{p{8cm} >{\raggedleft\arraybackslash}p{3cm} >{\raggedleft\arraybackslash}p{3cm} >{\raggedleft\arraybackslash}p{2cm}}
        \toprule
        & \multicolumn{1}{c}{\textbf{Control}} & \multicolumn{1}{c}{\textbf{SPRINT}} \\
        \multicolumn{1}{c}{\multirow{-2}{*}{\textbf{Characteristic}}} & \multicolumn{1}{c}{($n=961$)} & \multicolumn{1}{c}{($n=1002$)} & \multicolumn{1}{c}{\multirow{-2}{*}{\textbf{$p$-value}}} \\
        \midrule
        
        
        
        Baseline systolic BP, mmHg (Mean (SD)) 
          & 139.27 (16.05) & 138.65 (15.60) & 0.379 \\
        
        Baseline diastolic BP, mmHg (Mean (SD)) 
          & 74.67 (12.14) & 75.28 (12.14) & 0.267 \\
        
        Body-mass index, kg/m$^2$ (Mean (SD)) 
          & 29.35 (5.53) & 29.49 (5.74) & 0.586 \\
        
        Number of antihypertensive agents (Mean (SD)) 
          & 2.12 (1.01) & 2.06 (1.00) & 0.165 \\
        
        Framingham 10-year CVD risk, \% (Mean (SD)) 
          & 21.66 (11.71) & 21.17 (11.56) & 0.352 \\
        
        Framingham 10-year CVD risk score $\ge 15\%$ 
          & 634 (66.0\%) & 656 (65.5\%) & 0.814\\
        
        Statin use: Yes 
          & 519 (54.0\%) & 512 (51.1\%) & 0.197\\
        
        Daily aspirin use: Yes 
          & 530 (55.2\%) & 575 (57.4\%) & 0.318\\
        
        Daily NSAID use: Yes 
          & 114 (11.9\%) & 132 (13.2\%) &0.380 \\
        
        \bottomrule
        \end{tabular}
        }
    \end{center}
\end{table}

\newpage

\begin{table}[htbp]
    \begin{center}
        \caption{\label{appendix:table:sprint_comorb}Comorbidities}
        \resizebox{\textwidth}{!}{
        \begin{tabular}{p{8cm} >{\raggedleft\arraybackslash}p{3cm} >{\raggedleft\arraybackslash}p{3cm} >{\raggedleft\arraybackslash}p{2cm}}
        \toprule
        & \multicolumn{1}{c}{\textbf{Control}} & \multicolumn{1}{c}{\textbf{SPRINT}} \\
        \multicolumn{1}{c}{\multirow{-2}{*}{\textbf{Comorbidity}}} & \multicolumn{1}{c}{($n=961$)} & \multicolumn{1}{c}{($n=1002$)} & \multicolumn{1}{c}{\multirow{-2}{*}{\textbf{$p$-value}}} \\
        \midrule
        
        History of cardiovascular disease   & 240 (25.0\%) & 251 (25.0\%) & 0.969\\[0.3em]
        
        
        Atrial fibrillation/atrial flutter  & 94 (9.8\%) & 120 (12.0\%) &0.119  \\
        
        Coronary artery disease   & 151 (15.7\%) & 179 (17.9\%) & 0.203 \\
        
        Heart attack  & 88 (9.2\%) & 88 (8.8\%) & 0.771\\
        
        Congestive heart failure  & 57 (5.9\%) & 57 (5.7\%) & 0.818\\
        
        Irregular heart beat  & 177 (18.4\%) & 199 (19.9\%) & 0.417 \\
        
        Gastric or peptic ulcer  & 105 (10.9\%) & 108 (10.8\%) & 0.916 \\
        
        Crohn’s disease, ulcerative colitis, or IBD   & 27 (2.8\%) & 33 (3.3\%) & 0.534 \\
        
        Diverticulitis  & 134 (13.9\%) & 140 (14.0\%) & 0.986 \\
        
        Chronic hepatitis or liver cirrhosis  & 19 (2.0\%) & 18 (1.8\%) & 0.769 \\
        
        Gallbladder disease or gallstones  & 110 (11.4\%) & 130 (13.0\%) & 0.302\\
        
        Kidney or bladder infections   & 192 (20.0\%) & 214 (21.4\%) & 0.451 \\
        
        Benign prostatic hypertrophy, BPH  & 229 (23.8\%) & 235 (23.5\%) &0.844  \\
        
        Prostatitis & 49 (5.1\%) & 63 (6.3\%) & 0.256\\
        
        Osteoarthritis or degenerative arthritis  & 319 (33.2\%) & 320 (31.9\%) &0.552 \\
        
         Rheumatoid arthritis & 84 (8.7\%) & 95 (9.5\%) &  0.569\\
        
        Gout & 209 (21.7\%) & 196 (19.6\%) & 0.231\\
        
        Any other type of arthritis & 208 (21.6\%) & 229 (22.9\%) & 0.519\\
        
        Hip problems & 155 (16.1\%) & 168 (16.8\%) & 0.703 \\
        
        Cancer (not including skin cancer)  & 153 (15.9\%) & 177 (17.7\%) & 0.302\\
        
        Skin cancer   & 185 (19.3\%) & 177 (17.7\%) &0.365  \\
        
        Peripheral vascular disease (PVD)   & 80 (8.3\%) & 73 (7.3\%) &0.391 \\
        
        Seizure disorder   & 13 (1.4\%) & 6 (0.6\%) &  0.088\\
        
        Stroke & 7 (0.7\%) & 6 (0.6\%) & 0.723\\
        
         Transient ischemic attack (TIA)  & 36 (3.7\%) & 31 (3.1\%) & 0.426\\
        
        Thyroid disease   & 135 (14.0\%) & 145 (14.5\%) &0.789  \\
        
        Anemia or low blood count & 179 (18.6\%) & 176 (17.6\%) & 0.541\\
        
        Diabetes or high blood sugar  & 21 (2.2\%) & 24 (2.4\%) & 0.756\\
        
        Hypertension or high blood pressure 
          & 903 (94.0\%) & 947 (94.5\%) & 0.603\\
        
        Low back pain   & 429 (44.6\%) & 482 (48.1\%) & 0.124\\
        
        \bottomrule
        \end{tabular}
        }
    \end{center}
\end{table}

\newpage

\addtocounter{table}{-1}
\begin{table}[htbp]
    \begin{center}
        \caption{Comorbidities (continued)}
        \resizebox{\textwidth}{!}{
        \begin{tabular}{p{8cm} >{\raggedleft\arraybackslash}p{3cm} >{\raggedleft\arraybackslash}p{3cm} >{\raggedleft\arraybackslash}p{2cm}}
        \toprule
        & \multicolumn{1}{c}{\textbf{Control}} & \multicolumn{1}{c}{\textbf{SPRINT}} \\
        \multicolumn{1}{c}{\multirow{-2}{*}{\textbf{Comorbidity}}} & \multicolumn{1}{c}{($n=961$)} & \multicolumn{1}{c}{($n=1002$)} & \multicolumn{1}{c}{\multirow{-2}{*}{\textbf{$p$-value}}} \\
        \midrule
        
        Cataracts & 388 (40.4\%) & 406 (40.5\%) & 0.948 \\
        
        Schizophrenia 
          & 5 (0.5\%) & 4 (0.4\%) &0.691 \\
        
        Depression  & 179 (18.6\%) & 157 (15.7\%) &0.082 \\
        
        Bipolar or manic depressive disorder 
          & 11 (1.1\%) & 20 (2.0\%) &0.130 \\
        
        Anxiety or panic disorder  & 86 (8.9\%) & 80 (8.0\%) & 0.442\\
        
        Post-traumatic stress disorder (PTSD)  & 33 (3.4\%) & 35 (3.5\%) &0.943 \\
        
        Alcohol abuse  & 37 (3.9\%) & 31 (3.1\%) & 0.360\\
        
        \bottomrule
        \end{tabular}
        }
    \end{center}
\end{table}

\clearpage

\begin{table}[htbp]
    \begin{center}
        \caption{\label{appendix:table:sprint_lab}Laboratory measures}
        \resizebox{\textwidth}{!}{
        \begin{tabular}{p{8cm} >{\raggedleft\arraybackslash}p{3cm} >{\raggedleft\arraybackslash}p{3cm} >{\raggedleft\arraybackslash}p{2cm}}
        \toprule
        & \multicolumn{1}{c}{\textbf{Control}} & \multicolumn{1}{c}{\textbf{SPRINT}} \\
        \multicolumn{1}{c}{\multirow{-2}{*}{\textbf{Laboratory measure}}} & \multicolumn{1}{c}{($n=961$)} & \multicolumn{1}{c}{($n=1002$)} & \multicolumn{1}{c}{\multirow{-2}{*}{\textbf{$p$-value}}} \\
        \midrule
        
        Serum creatinine, mg/dL (Mean (SD)) & 1.43 (0.37) & 1.45 (0.40) & 0.514 \\
        
        Estimated GFR, ml/min/1.73 m$^2$ (Mean (SD)) & 47.87 (9.41) & 47.73 (9.59) & 0.748 \\
        
        log(UMALCR) at baseline (Mean (SD)) & 2.96 (1.43) & 2.97 (1.42) & 0.868 \\
        
        
        Fasting total cholesterol, mg/dL (Mean (SD)) & 183.34 (39.07) & 185.56 (40.89) & 0.219 \\
        
        Fasting HDL cholesterol, mg/dL (Mean (SD)) & 51.92 (14.63) & 52.68 (14.92) & 0.257 \\
        
        Fasting total triglycerides, mg/dL (Mean (SD)) & 128.91 (64.47) & 122.93 (59.14) & 0.032 \\
        
        Fasting plasma glucose, mg/dL (Mean (SD)) & 98.19 (11.65) & 98.18 (13.37) & 0.994 \\
        
        Urea nitrogen, mg/dL (Mean (SD)) & 24.54 (7.94) & 24.30 (8.37) & 0.516 \\
        
        Chloride, mmol/L (Mean (SD)) & 103.36 (2.92) & 103.48 (2.98) & 0.364 \\
        
        Bicarbonate, mmol/L (Mean (SD)) & 25.66 (2.68) & 25.65 (2.78) & 0.935 \\
        
        Urine creatinine, mg/dL (Mean (SD)) & 128.91 (69.94) & 131.24 (79.54) & 0.491 \\
        
        Potassium, mmol/L (Mean (SD)) & 4.32 (0.47) & 4.30 (0.48) & 0.508 \\
        
        LDL cholesterol, mg/dL (Mean (SD)) & 105.64 (32.88) & 108.29 (35.14) & 0.086 \\
        
        Sodium, mmol/L (Mean (SD)) & 140.34 (2.41) & 140.37 (2.47) & 0.821 \\
        
        Urine microalbumin, mg/L (Mean (SD)) & 80.11 (224.55) & 79.80 (270.10) & 0.978 \\
        
        \bottomrule
        \end{tabular}
        }
    \end{center}
\end{table}

\clearpage

\bibliographystyle{apalike}
\bibliography{Bibliography}


\end{document}